\documentclass[12pt]{article}
\usepackage{amsmath}
\usepackage[unicode,bookmarks,bookmarksopen,bookmarksopenlevel=2,colorlinks,linkcolor=blue,citecolor=green]{hyperref}

\input{tcilatex}

\begin{document}

\title{New Hamiltonian formalism and Lagrangian representations for
integrable hydrodynamic type systems.}
\author{M.V. Pavlov \\
%EndAName
Lebedev Physical Institute, Moscow}
\date{}
\maketitle

\begin{abstract}
New Hamiltonian formalism based on the theory of conjugate curvilinear
coordinate nets is established. All formulas are ``mirrored'' to
corresponding formulas in Hamiltonian formalism constructed by B.A. Dubrovin
and S.P. Novikov (in a flat case) and E.V. Ferapontov (in a non-flat case).
In the ``mirrored-flat'' case Lagrangian formulation is found.
Multi-Hamiltonian examples are presented. In particular Egorov's case,
generalizations of local Nutku--Olver's Hamiltonian structure and
corresponding Sheftel--Teshukov's recursion operator are presented. An 
\textit{infinite} number of \textit{local} Hamiltonian structures of \textit{%
all odd} orders is found.
\end{abstract}

\vspace{1cm}

\textit{In honour of Yavuz Nutku}

\vspace{1cm}

\tableofcontents

\textit{keywords}: Hamiltonian structure, Egorov metric, hydrodynamic type
system, Riemann invariant, Lagrangian representation.


\section{Introduction}

The modern theory of integrable hydrodynamic type systems --- i.e.\
quasilinear systems of first-order PDE's (see e.g. \textbf{\cite{Yanenko}}, 
\textbf{\cite{Serre}})%
\begin{equation}
\begin{array}{l}
\left( 
\begin{array}{c}
u_{t}^{1} \\ 
\vdots \\ 
u_{t}^{n}%
\end{array}%
\right) =\left( 
\begin{array}{ccc}
v_{1}^{1}(u) & \cdots & v_{n}^{1}(u) \\ 
\cdot & \cdots & \cdot \\ 
v_{1}^{n}(u) & \cdots & v_{n}^{n}(u)%
\end{array}%
\right) \left( 
\begin{array}{c}
u_{x}^{1} \\ 
\vdots \\ 
u_{x}^{n}%
\end{array}%
\right) ,%
\end{array}
\label{Ts:1+1h}
\end{equation}%
where $u^{i}=u^{i}(x,t)$, $i=1,\cdots ,n$ are the field variables (unknown
functions), $x$ and $t$ are one-dimensional space and time variables ---
started with the paper \textbf{\cite{Dubr+Nov}} where a general Hamiltonian
formalism for such systems was proposed. Physical examples of (\textbf{\ref%
{Ts:1+1h}}) with a ``natural'' Hamiltonian structure are numerous (see e.g.\ 
\textbf{\cite{bibTs:Bao}}, \textbf{\cite{bibTs:DzV}}, \textbf{\cite%
{bibTs:Hol}}, \textbf{\cite{bibTs:Katz}}) for physical examples of systems
with Hamiltonian structures originating from infinite-dimensional Lie
algebras naturally related to the respective systems; such Hamiltonian
structures are a particular case of the more general Hamiltonian structure
described by B.A. Dubrovin and S.P. Novikov. Their generalization allows us
to cast into the Hamiltonian form another vast class of hydrodynamic type
systems (\textbf{\ref{Ts:1+1h}}), derived from integrable nonlinear
equations via the so called ``averaging'' procedure, which gives a system
(usually of type (\textbf{\ref{Ts:1+1h}})) describing slowly varying
(``modulated'') quasiperiodic solutions (see e.g. \textbf{\cite{bibTs:CEM}}, 
\textbf{\cite{Dubr+Nov}}, \textbf{\cite{Erc}}, \textbf{\cite{FFML}}, \textbf{%
\cite{Krich+Whitham}}, \textbf{\cite{Lax}} \textbf{\cite{Tsar}}, \textbf{%
\cite{Tsar+geom}}). These Whitham equations possess the Hamiltonian
structure of the Dubrovin--Novikov type. Some particular cases of this
Hamiltonian structure for averaged equations were studied in \textbf{\cite%
{Hay}}.

This method may be applied to a wide range of ``averaged'' integrable
equations which possess alongside with the Hamiltonian structure of the
Dubrovin--Novikov type the remarkable property of \textit{diagonalizability}%
: after a suitable change of the field variables $r^{i}(\mathbf{u})$ in (%
\textbf{\ref{Ts:1+1h}}) one obtains the hydrodynamic type system%
\begin{equation}
r_{t}^{i}=v^{i}(\mathbf{r})r_{x}^{i}\text{, \ \ \ \ \ \ }i=1,2,...,N.
\label{rim}
\end{equation}%
(no summation over repeated indices hereafter!) with diagonal matrix $v^{i}(%
\mathbf{r})\delta _{j}^{i}$. These new variables $r^{i}$ are called \textit{%
Riemann invariants} (in the Whitham theory they are \textit{branch points}
of corresponding Riemann surfaces, see e.g. \textbf{\cite{FFML}}, \textbf{%
\cite{Maks+algebr}}). For $n=2$ \textit{every} hyperbolic system (\textbf{%
\ref{Ts:1+1h}}) may be diagonalized; for $n>2$ this is not true in general,
a special criterion \textbf{\cite{bibTs:Haan}} shall be applied to check
diagonalizability of (\textbf{\ref{Ts:1+1h}}).

On the other hand the methods of \textbf{\cite{Tsar}} (see also \textbf{\cite%
{Tsar+geom}}) lead to an unexpected link between the theory of Hamiltonian
diagonalizable hydrodynamic type systems and a classical object of local
differential geometry intensively studied in the end of 19th and beginning
of 20th century --- orthogonal curvilinear coordinates in the flat Euclidean
space $R^{n}$. Practically all ``physical'' entities (conservation laws,
symmetries etc.) have their counterparts in the theory of orthogonal
curvilinear coordinates; amazingly enough the basic formulas relating
conservation laws, symmetries and the Hamiltonian structure of integrable
systems of hydrodynamic type may be found in \textbf{\cite{bibTs:dar1}}!
Further investigation discovered a deep relation between other types of
integrable nonlinear systems studied in the modern theory of integrable
physical equations and classical problems studied in \textbf{\cite%
{bibTs:bianchi,bibTs:Boul,bibTs:dar1,bibTs:dar2,bibTs:demoul,bibTs:Eisenh,bibTs:Gui}%
} and other papers at the beginning of previous century.

In this paper we expose this remarkable link as well as some recently
discovered deeper relations between the classical differential geometry
flourished at the end of the 19th century and the modern theory of
integrable nonlinear PDEs appearing in different applications to
mathematical physics.

The theory of semi-Hamiltonian hydrodynamic type systems is a most developed
part of integrable PDE systems (see \textbf{\cite{Dubr+Nov}}, \textbf{\cite%
{Tsar}}, \textbf{\cite{Tsar+geom}}). A one of such reasons is that this
theory is based on classical differential geometry, and more precisely, on
the theory of conjugate curvilinear coordinate nets (see \textbf{\cite%
{bibTs:dar1}}). For instance, the local Hamiltonian formalism associated
with the differential-geometric Poisson brackets of the first order is
connected with theory of orthogonal curvilinear coordinate nets (see \textbf{%
\cite{Dubr}}, \textbf{\cite{Krich}}, \textbf{\cite{Tsar}}, \textbf{\cite%
{Tsar+geom}}, \textbf{\cite{Zakh}}).

This paper is devoted to an alternative construction based on so-called 
\textit{anti-flatness property}. An existence and, moreover, an explicit
construction of Hamiltonian structures play important role in integrability
of hydrodynamic type systems, because a general solution of semi-Hamiltonian
hydrodynamic type system depends on general solution of a linear problems
describing conservation laws and commuting flows, which are related by the
Hamiltonian structure.

In general case the problem is following: it is necessary to describe all
possible Hamiltonian formalisms associated with integrable hydrodynamic type
systems or with conjugate curvilinear coordinate nets. The local Hamiltonian
formalism determined by a differential operator of the first order is
constructed in (\textbf{\cite{Dubr+Nov}} (see also \textbf{\cite{Tsar}}, 
\textbf{\cite{Tsar+geom}}). The nonlocal Hamiltonian formalism of the first
order associated with a flat normal bundle is constructed in \textbf{\cite%
{Fer+trans}}. This paper is devoted to the nonlocal Hamiltonian formalism
associated with the local Lagrangian representation of corresponding
hydrodynamic type systems. Suppose a given hydrodynamic type system
possesses the local Hamiltonian structure of the Dubrovin--Novikov type (see 
\textbf{\cite{Dubr+Nov}}) and the above mentioned Lagrangian representation,
simultaneously. It means, that this hydrodynamic type system has local
Hamiltonian and local symplectic structures. A most amazing consequence is
that the corresponding hydrodynamic type system has infinitely many local
Hamiltonian structures of all odd orders (as well as infinitely many local
symplectic structures and corresponding local Lagrangian representations).

This paper is organized as follows. In Section \textbf{2}, the relationship
between local Hamiltonian formalism of the Dubrovin--Novikov type and
orthogonal curvilinear coordinate nets is briefly described. In Section 
\textbf{3}, the \textit{first} ``puzzle'' hidden in the paper \textbf{\cite%
{NO}} written by Yavuz Nutku and Peter Olver is formulated. In Section 
\textbf{4}, integrable hydrodynamic type systems determined by a special
local Lagrangian representation are found. The existence of such Lagrangians
is based on the concept ``anti-flatness'', which is established in this
paper. In Section \textbf{5}, simultaneously, hydrodynamic type systems
associated with a local Hamiltonian structure and with above mentioned local
Lagrangian representation are considered. Local Hamiltonian structures of
all odd orders are described. In Section \textbf{6}, hydrodynamic type
systems possessing \textit{anti-flat} multi-Hamiltonian structures are
discussed. In the Section \textbf{7}, the above construction is restricted
on the symmetric case. \textbf{Statement}: suppose the Egorov hydrodynamic
type system has just one known local Hamiltonian structure of the
Dubrovin--Novikov type (or vise versa has just one known local Lagrangian
representation), then \textit{infinitely many} other \textit{local}
Hamiltonian structures of \textit{all odd} orders can be constructed. All of
them can be expressed via corresponding solutions of the WDVV equation (see 
\textbf{\cite{Dubr}}, \textbf{\cite{Krich+wdvv}}, \textbf{\cite{Mokh}}, 
\textbf{\cite{Maks+wdvv}}). In Section \textbf{8}, the \textit{second}
``puzzle'' hidden in the paper \textbf{\cite{NO}} written by Yavuz Nutku and
Peter Olver is formulated. A natural interpretation via local Hamiltonian
structures of the Dubrovin--Novikov type is given. In Section \textbf{9}, a
generalization of above construction on a nonlocal case associated with
surfaces with a flat normal bundle is briefly discussed. In Section \textbf{%
10}, the above local Lagrangian formulation is extended on higher order
commuting flows of hydrodynamic type systems.

\section{Flat case}

The theory of \textit{\textbf{conjugate} curvilinear coordinate nets} (see 
\textbf{\cite{bibTs:dar1}}) is based on nonlinear PDE system%
\begin{equation}
\partial _{i}\beta _{jk}=\beta _{ji}\beta _{ik}\text{, \ \ \ \ \ \ \ }i\neq
j\neq k,  \label{non}
\end{equation}%
where \textit{rotation coefficients} $\beta _{ik}$ are functions of $N$
independent variables $r^{k}$ (which are called \textit{principal curvature
coordinates}, see \textbf{\cite{Terng}}). A general solution of this system
is parameterized by $N(N-1)/2$ functions of two variables. The above
nonlinear PDE system is a consequence of the compatibility conditions $%
\partial _{i}(\partial _{k}H_{j})=\partial _{k}(\partial _{i}H_{j})$,\ $%
\partial _{i}(\partial _{k}\psi _{j})=\partial _{k}(\partial _{i}\psi _{j})$%
, $i\neq j\neq k$, where \textit{Lame coefficients} $H_{j}$ and \textit{%
adjoint} Lame coefficients $\psi _{j}$ are solutions of a linear PDE system
and an \textit{adjoint} linear PDE system, respectively%
\begin{equation}
\partial _{i}H_{k}=\beta _{ik}H_{i}\text{, \ \ \ \ \ \ \ \ }\partial
_{k}\psi _{i}=\beta _{ik}\psi _{k}\text{, \ \ \ \ \ \ \ }i\neq k,
\label{lin}
\end{equation}%
which have general solutions parameterized by $N$ arbitrary functions of a
single variable for any given rotation coefficients $\beta _{ik}$ satisfying
(\textbf{\ref{non}}). In general case a relationship between these linear
PDE systems is unknown.

A link connecting conjugate curvilinear coordinate nets and integrable
hydrodynamic type systems is following.

\begin{itemize}
\item Let $\bar{H}_{i}$ and $\tilde{H}_{i}$ be two particular solutions of
the first linear PDE system (\textbf{\ref{lin}}).

\item Let field variables $r^{k}$ be functions of two independent variables $%
x$ and $t$.

\item Let construct a hydrodynamic type system written via Riemann
invariants $r^{k}$%
\begin{equation}
r_{t}^{i}=\frac{\tilde{H}_{i}}{\bar{H}_{i}}r_{x}^{i}.  \label{first}
\end{equation}

\item Let field variables $r^{k}$ be functions of three independent
variables $x$, $t$ and $\tau $, simultaneously. Then the above hydrodynamic
type system possesses a commuting flow (i.e. $(r_{\tau
}^{i})_{t}=(r_{t}^{i})_{\tau }$)%
\begin{equation}
r_{\tau }^{i}=\frac{H_{i}}{\bar{H}_{i}}r_{x}^{i},  \label{sec}
\end{equation}%
parameterized by $N$ arbitrary functions of a single variable (see (\textbf{%
\ref{lin}}) and \textbf{\cite{Tsar}}, \textbf{\cite{Tsar+geom}}).

\item Let introduce the function $h$ such that%
\begin{equation}
\partial _{i}h=\psi _{i}\bar{H}_{i}.  \label{zak}
\end{equation}%
Then the above hydrodynamic type system has a conservation law (written in
the potential form)%
\begin{equation*}
d\xi =hdx+gdt+fd\tau ,
\end{equation*}%
where%
\begin{equation*}
\partial _{i}g=\psi _{i}\tilde{H}_{i}\text{, \ \ \ \ \ \ \ \ }\partial
_{i}f=\psi _{i}H_{i}.
\end{equation*}%
Thus, these hydrodynamic type systems possess infinitely many conservation
laws parameterized by $N$ arbitrary functions of a single variable.

\item A general solution of the hydrodynamic type system (\textbf{\ref{lin}}%
) is given (in an implicit form) by the generalized hodograph method (see 
\textbf{\cite{Tsar}})%
\begin{equation*}
x\bar{H}_{i}+t\tilde{H}_{i}=H_{i}.
\end{equation*}
\end{itemize}

The first problem in the theory of hydrodynamic type systems (\textbf{\ref%
{first}}) is a description of solutions of the nonlinear PDE system (\textbf{%
\ref{non}}), the second problem is a description of solutions of the linear
PDE systems with variable coefficients (\textbf{\ref{lin}}). These problems
are very complicated in general. By this reason, conjugate curvilinear
coordinate nets can be separated on some sub-classes due to some criterion
(features, characteristics etc...), then corresponding conjugate curvilinear
coordinate nets can be integrated within corresponding sub-classes.

The approach developed by B.A. Dubrovin, S.P. Novikov (see \textbf{\cite%
{Dubr+Nov}}) and S.P. Tsarev (see \textbf{\cite{Tsar}}) is based on a local
Hamiltonian formalism for hydrodynamic type systems (\textbf{\ref{first}})
connected with the theory of \textit{\textbf{orthogonal} curvilinear
coordinate nets} (see \textbf{\cite{Tsar}}, \textbf{\cite{Tsar+geom}}),
which is a sub-class of the theory of conjugate curvilinear coordinate nets.

\begin{itemize}
\item Let establish a link of the first order between solutions of the
linear and adjoint linear PDE systems (\textbf{\ref{lin}})%
\begin{equation}
H_{i}=\partial _{i}\psi _{i}+\underset{m\neq i}{\sum }\beta _{mi}\psi _{m}.
\label{lok}
\end{equation}

\item Then the \textit{flatness} condition is given by (see \textbf{\cite%
{bibTs:dar1}})%
\begin{equation}
\partial _{i}\beta _{ik}+\partial _{k}\beta _{ki}+\underset{m\neq i,k}{\sum }%
\beta _{mi}\beta _{mk}=0\text{, \ \ \ \ \ \ }i\neq k  \label{flat}
\end{equation}%
which can be obtained by the substitution (\textbf{\ref{lok}}) in (\textbf{%
\ref{lin}}). Thus, the nonlinear PDE system (\textbf{\ref{non}}) and (%
\textbf{\ref{flat}}) is a consequence of compatibility conditions of the
linear PDE system (\textbf{\ref{lin}}) and (\textbf{\ref{lok}}). It means
that orthogonal curvilinear coordinate nets are described by the nonlinear
PDE system (\textbf{\ref{non}}) and (\textbf{\ref{flat}}).

\item If l.h.s. of (\textbf{\ref{lok}}) is vanished, then $N$ particular
solutions $\bar{\psi}_{(\gamma )i}$ are determined by the linear ODE systems
(for each \textit{fixed} index $i$; see (\textbf{\ref{lin}}) and (\textbf{%
\ref{lok}}))%
\begin{equation*}
\partial _{i}\bar{\psi}_{(\gamma )k}=\beta _{ki}\bar{\psi}_{(\gamma )i}\text{%
, \ \ \ \ \ \ }i\neq k\text{, \ \ \ \ \ \ }\partial _{i}\bar{\psi}_{(\gamma
)i}+\underset{m\neq i}{\sum }\beta _{mi}\bar{\psi}_{(\gamma )m}=0.
\end{equation*}

\item These linear ODE systems have $N(N+1)/2$ constraints (``first
integrals'')%
\begin{equation}
\bar{g}_{\beta \gamma }=\sum \bar{\psi}_{(\beta )m}\bar{\psi}_{(\gamma )m}=%
\limfunc{const}.  \label{metr}
\end{equation}%
This is non-degenerate matrix. Then $\bar{\psi}_{i}^{(\beta )}=\bar{g}%
^{\beta \gamma }\bar{\psi}_{(\gamma )i}$.

\item Let $a^{\gamma }$ be flat coordinates, i.e. $\partial _{i}a^{\gamma }=%
\bar{\psi}_{i}^{(\gamma )}\bar{H}_{i}$ (see (\textbf{\ref{zak}})) and%
\begin{equation}
\frac{\partial r^{i}}{\partial a^{\gamma }}=\frac{\bar{\psi}_{(\gamma )i}}{%
\bar{H}_{i}}.  \label{invert}
\end{equation}%
Then (\textbf{\ref{first}}) under a point transformation $r^{i}=r^{i}(%
\mathbf{a})$ can be written in the conservative form%
\begin{equation}
a_{t}^{\beta }=\bar{g}^{\beta \gamma }D_{x}\frac{\partial \tilde{h}}{%
\partial a^{\gamma }},  \label{dn}
\end{equation}%
where $\tilde{h}(\mathbf{a})$ is a Hamiltonian density, the momentum density 
$\bar{h}(\mathbf{a})$ is a \textit{quadratic} expression with respect to
flat coordinates $a^{\gamma }$%
\begin{equation}
\bar{h}=\frac{1}{2}\bar{g}_{\beta \gamma }a^{\beta }a^{\gamma }.  \label{imp}
\end{equation}%
It means (see \textbf{\cite{Maks+Tsar}}), that the above hydrodynamic type
system has two extra conservation laws (of the momentum and of the energy,
respectively) associated with the above local Hamiltonian structure%
\begin{equation*}
\partial _{t}\bar{h}=D_{x}\left( a^{\beta }\frac{\partial \tilde{h}}{%
\partial a^{\beta }}-\tilde{h}\right) \text{, \ \ \ \ \ \ \ \ \ }\partial
_{t}\tilde{h}=D_{x}\left( \bar{g}^{\beta \gamma }\frac{\partial \tilde{h}}{%
\partial a^{\beta }}\frac{\partial \tilde{h}}{\partial a^{\gamma }}\right) .
\end{equation*}%
Then the above conservation law densities $\bar{h}$ and $\tilde{h}$ can be
expressed via Riemann invariants $r^{k}$ (in quadratures)%
\begin{eqnarray*}
\partial _{i}\bar{h} &=&\bar{\psi}_{i}\bar{H}_{i}\text{,\ \ \ \ \ \ \ \ \ \ }%
\partial _{i}\left( a^{\beta }\frac{\partial \tilde{h}}{\partial a^{\beta }}-%
\tilde{h}\right) =\bar{\psi}_{i}\tilde{H}_{i}\text{, \ \ \ \ \ \ \ \ \ }%
\partial _{i}\frac{\partial \bar{h}}{\partial a^{\beta }}=\bar{\psi}_{(\beta
)i}\bar{H}_{i}, \\
&& \\
\partial _{i}\tilde{h} &=&\tilde{\psi}_{i}\bar{H}_{i}\text{, \ \ \ \ \ \ \ \ 
}\partial _{i}\left( \bar{g}^{\beta \gamma }\frac{\partial \tilde{h}}{%
\partial a^{\beta }}\frac{\partial \tilde{h}}{\partial a^{\gamma }}\right) =%
\tilde{\psi}_{i}\tilde{H}_{i}\text{, \ \ \ \ \ \ \ }\partial _{i}\frac{%
\partial \tilde{h}}{\partial a^{\beta }}=\bar{\psi}_{(\beta )i}\tilde{H}_{i},
\end{eqnarray*}%
where (see (\textbf{\ref{lok}}))%
\begin{equation}
\bar{H}_{i}=\partial _{i}\bar{\psi}_{i}+\underset{m\neq i}{\sum }\beta _{mi}%
\bar{\psi}_{m}\text{, \ \ \ \ \ \ \ \ }\tilde{H}_{i}=\partial _{i}\tilde{\psi%
}_{i}+\underset{m\neq i}{\sum }\beta _{mi}\tilde{\psi}_{m}\text{.}
\label{odin}
\end{equation}%
\textbf{Corollary}: The momentum density (see the first above equation) is
given by (cf. (\textbf{\ref{imp}}))%
\begin{equation}
\bar{h}=\frac{1}{2}\sum \bar{\psi}_{m}^{2}.  \label{mom}
\end{equation}

\item It means, that the \textit{\textbf{local} Poisson bracket of the
Dubrovin--Novikov type }in flat coordinates is given by%
\begin{equation*}
\{a^{\beta }(x),a^{\gamma }(x^{\prime })\}=\bar{g}^{\beta \gamma }\delta
^{\prime }(x-x^{\prime });
\end{equation*}%
in Riemann invariants%
\begin{equation*}
\{r^{i}(x),r^{j}(x^{\prime })\}=\hat{A}^{ij}\delta (x-x^{\prime }),
\end{equation*}%
where%
\begin{equation}
\hat{A}^{ii}=\frac{1}{\bar{H}_{i}}\partial _{x}\frac{1}{\bar{H}_{i}}\text{,
\ \ \ \ \ \ \ \ }\hat{A}^{ik}|_{k\neq i}=\frac{1}{\bar{H}_{i}\bar{H}_{k}}%
(\beta _{ki}r_{x}^{i}-\beta _{ik}r_{x}^{k}).  \label{lp}
\end{equation}

\item Thus, an arbitrary commuting flow (\textbf{\ref{sec}}) also can be
written in the same local canonical Hamiltonian form%
\begin{equation*}
a_{\tau }^{\beta }=\bar{g}^{\beta \gamma }D_{x}\frac{\partial h}{\partial
a^{\gamma }},
\end{equation*}%
where%
\begin{equation*}
\partial _{i}\left( a^{\beta }\frac{\partial h}{\partial a^{\beta }}%
-h\right) =\bar{\psi}_{i}H_{i}\text{,\ \ \ }\partial _{i}h=\psi _{i}\bar{H}%
_{i}\text{, \ \ }\partial _{i}\frac{\partial h}{\partial a^{\beta }}=\bar{%
\psi}_{(\beta )i}H_{i}\text{,\ \ \ }\partial _{i}\left( \bar{g}^{\beta
\gamma }\frac{\partial h}{\partial a^{\beta }}\frac{\partial h}{\partial
a^{\gamma }}\right) =\psi _{i}H_{i},
\end{equation*}%
and solutions $\psi _{i}$ and $H_{i}$ of linear PDE systems (\textbf{\ref%
{lin}}) are related by (\textbf{\ref{lok}}).

\item The relationship of the first order between the linear problems (%
\textbf{\ref{lin}}) is given by (\textbf{\ref{lok}}). Taking into account
another identity%
\begin{equation}
\frac{\partial h}{\partial a^{\beta }}=\sum \bar{\psi}_{(\beta )m}\psi _{m}%
\text{,}  \label{zuk}
\end{equation}%
an \textit{inverse} (nonlocal) relationship between the linear problems (%
\textbf{\ref{lin}}) is given by%
\begin{equation}
\psi _{i}=\bar{\psi}_{i}^{(\beta )}\frac{\partial h}{\partial a^{\beta }}.
\label{inv}
\end{equation}%
Indeed (see (\textbf{\ref{lok}})),%
\begin{equation*}
H_{i}=\partial _{i}\left( \bar{\psi}_{i}^{(\beta )}\frac{\partial h}{%
\partial a^{\beta }}\right) +\underset{m\neq i}{\sum }\beta _{mi}\bar{\psi}%
_{m}^{(\beta )}\frac{\partial h}{\partial a^{\beta }}\equiv \bar{\psi}%
_{i}^{(\beta )}\partial _{i}\frac{\partial h}{\partial a^{\beta }}\text{ \ \
\ \ }\Leftrightarrow \text{ \ \ \ \ }\partial _{i}\frac{\partial h}{\partial
a^{\beta }}=\bar{\psi}_{(\beta )i}H_{i}.
\end{equation*}
\end{itemize}

\section{\textit{First} Nutku--Olver's ``puzzle''}

The ideal gas dynamic system (see e.g. \textbf{\cite{Nutku}}, \textbf{\cite%
{NO}})%
\begin{equation}
\rho _{t}=(\rho u)_{x}\text{, \ \ \ \ }u_{t}=\left( \frac{u^{2}}{2}+\rho
f^{\prime \prime }(\rho )-f^{\prime }(\rho )\right) _{x}  \label{ideal}
\end{equation}%
has \textit{\textbf{only one local}} Hamiltonian structure of the \textit{%
Dubrovin--Novikov} type%
\begin{equation*}
\left( 
\begin{array}{c}
\rho  \\ 
u%
\end{array}%
\right) _{t}=\left( 
\begin{array}{cc}
0 & D_{x} \\ 
D_{x} & 0%
\end{array}%
\right) \left( 
\begin{array}{c}
\frac{\delta \mathbf{H}_{4}}{\delta \rho } \\ 
\frac{\delta \mathbf{H}_{4}}{\delta u}%
\end{array}%
\right) ,
\end{equation*}%
where the Hamiltonian and other lower functionals are given by%
\begin{equation*}
\mathbf{H}_{4}=\int \left[ \frac{\rho u^{2}}{2}+\rho f^{\prime }(\rho
)-2f(\rho )\right] dx\text{, \ \ \ \ }\mathbf{H}_{3}=\int \rho udx\text{, \
\ \ \ }\mathbf{H}_{2}=\int \rho dx\text{, \ \ \ \ }\mathbf{H}_{1}=\int udx%
\text{.}
\end{equation*}%
However, Ya. Nutku and P. Olver (see \textbf{\cite{NO}}) found an {\textbf{%
absolutely new local}} Hamiltonian structure determined by the {\textit{%
third order}} purely differential operator%
\begin{equation}
\hat{B}=\hat{R}^{2}\left( 
\begin{array}{cc}
0 & D_{x} \\ 
D_{x} & 0%
\end{array}%
\right) ,  \label{no}
\end{equation}%
where the recursion operator $\hat{R}$ is a purely differential operator of
the {\textit{first}} order%
\begin{equation}
\hat{R}=D_{x}(W_{x})^{-1}  \label{rek}
\end{equation}%
and the matrix $W$ is given by%
\begin{equation}
W=\left( 
\begin{array}{cc}
u & \rho  \\ 
f^{\prime \prime }(\rho ) & u%
\end{array}%
\right) .  \label{you}
\end{equation}

This paper is devoted to an investigation of this phenomenon. Indeed, this
local Hamiltonian structure of the third order has clear pure
differential-geometric meaning. Moreover, a corresponding
differential-geometric construction is generalized on $N$ component
hydrodynamic type systems.

So, we have the set of questions:

\begin{itemize}
\item \textbf{How to \textit{explain} the {\textit{ORIGIN}} of this local
Hamiltonian structure?}

\item \textbf{How to \textit{generalize} this local Hamiltonian structure on 
$N$ component case?}

\item {\ \textbf{How to \textit{recognize} that a given hydrodynamic type
system possesses such a local Hamiltonian structure?}}

\item \textbf{How to \textit{construct} such a local Hamiltonian structure?}

\item \textbf{How \textit{many} such local Hamiltonian structures a given
hydrodynamic type system can possess?}

\item \textbf{Is it any \textit{relationship} between these local
Hamiltonian structures and local Hamiltonian structures of the
Dubrovin--Novikov type?}
\end{itemize}

The key idea used for opening this puzzle is hidden in the local Lagrangian
representation for this hydrodynamic type system (see \textbf{\cite{Lagrang}}%
)%
\begin{equation}
S=\int \left[ \frac{1}{2}\frac{\rho _{x}u_{t}-u_{x}\rho _{t}}{%
u_{x}^{2}-f^{\prime \prime \prime }(\rho )\rho _{x}^{2}}-\rho \right] dxdt.
\label{lag}
\end{equation}

\section{Local Lagrangian representations}

The Lagrangian%
\begin{equation*}
S=\int [g_{k}(r,r_{x},r_{xx},...)r_{t}^{k}-h(r,r_{x},r_{xx},...)]dxdt
\end{equation*}%
determines the Euler--Lagrange equations%
\begin{equation*}
\hat{M}_{ik}r_{t}^{k}=\frac{\delta \mathbf{H}}{\delta r^{i}}
\end{equation*}%
in the Hamiltonian form (see \textbf{\cite{Mokhov}})%
\begin{equation*}
r_{t}^{i}=\hat{K}^{ij}\frac{\delta \mathbf{H}}{\delta r^{j}},
\end{equation*}%
where the Hamiltonian $\mathbf{H}=\int h(r,r_{x},r_{xx},...)dx$, and the
Hamiltonian operator $\hat{K}^{ij}$ is an inverse operator to the symplectic
operator%
\begin{equation}
\hat{M}_{ij}=\frac{\partial g_{i}}{\partial r_{(n)}^{j}}%
D_{x}^{n}-(-1)^{n}D_{x}^{n}\frac{\partial g_{j}}{\partial r_{(n)}^{i}}.
\label{sym}
\end{equation}%
Let us restrict our further consideration on the case%
\begin{equation*}
S=\int [g_{k}(r,r_{x})r_{t}^{k}-h(r,r_{x},r_{xx},...)]dxdt.
\end{equation*}%
Then corresponding Euler--Lagrange equations can be written in the form%
\begin{equation*}
\hat{M}_{ik}r_{t}^{k}=\frac{\delta \mathbf{H}}{\delta r^{i}},
\end{equation*}%
where a corresponding symplectic operator is given by (see (\textbf{\ref{sym}%
}))%
\begin{equation*}
\hat{M}_{ik}=\frac{\partial g_{k}}{\partial r^{i}}-\frac{\partial g_{i}}{%
\partial r^{k}}-\left( \frac{\partial g_{i}}{\partial r_{x}^{k}}D_{x}+D_{x}%
\frac{\partial g_{k}}{\partial r_{x}^{i}}\right) .
\end{equation*}

If we choose%
\begin{equation*}
g_{k}(r,r_{x})=\frac{\bar{H}_{k}^{2}(\mathbf{r})}{2r_{x}^{k}},
\end{equation*}%
then corresponding components of the above symplectic operator are given by
(cf. (\textbf{\ref{lp}}))%
\begin{equation}
\hat{M}_{ii}=\frac{\bar{H}_{i}}{r_{x}^{i}}D_{x}\frac{\bar{H}_{i}}{r_{x}^{i}}%
\text{, \ \ \ \ \ }\hat{M}_{ik}|_{k\neq i}=\bar{H}_{i}\bar{H}_{k}\left( 
\frac{\beta _{ik}}{r_{x}^{k}}-\frac{\beta _{ki}}{r_{x}^{i}}\right) ,
\label{lsym}
\end{equation}%
where%
\begin{equation}
\beta _{ik}=\frac{\partial _{i}\bar{H}_{k}}{\bar{H}_{i}}\text{, \ \ \ \ }%
k\neq i.  \label{rot}
\end{equation}

\textbf{Theorem 1}: \textit{The Hamiltonian operator}%
\begin{equation}
\hat{K}^{ij}=\bar{\varepsilon}^{\alpha \beta }w_{(\alpha
)}^{i}r_{x}^{i}D_{x}^{-1}w_{(\beta )}^{j}r_{x}^{j}  \label{nham}
\end{equation}%
\textit{is an \textbf{inverse} operator to the above symplectic operator} $%
\hat{M}_{ik}$ \textit{iff}

\textbf{1}. $\beta _{ik}$ \textit{are rotation coefficients of a
corresponding conjugate curvilinear coordinate net determined by the
Bianchi--Darboux--Lame system }(\textbf{\ref{non}}) \textit{and by the {%
``anti-flatness''} condition}%
\begin{equation}
\partial _{i}\beta _{ki}+\partial _{k}\beta _{ik}+\underset{m\neq i}{\sum }%
\beta _{im}\beta _{km}=0\text{, \ \ \ \ }i\neq k;  \label{antiflat}
\end{equation}

\textbf{2}. \textit{affinors }$w_{(\alpha )}^{i}$\textit{\ determine }$N$ 
\textit{commuting hydrodynamic type systems (structural flows)}%
\begin{equation}
r_{t^{\alpha }}^{i}=w_{(\alpha )}^{i}r_{x}^{i},  \label{struc}
\end{equation}%
\textit{where }$w_{(\alpha )}^{i}=\bar{H}_{(\alpha )i}/\bar{H}_{i}$, \textit{%
and }$\bar{H}_{(\alpha )i}$ \textit{are solutions of the linear ODE systems}%
\begin{equation*}
\partial _{i}\bar{H}_{(\alpha )k}=\beta _{ik}\bar{H}_{(\alpha )i}\text{, \ \
\ \ }i\neq k\text{, \ \ \ \ \ }\partial _{i}\bar{H}_{(\alpha )i}+\underset{%
m\neq i}{\sum }\beta _{im}\bar{H}_{(\alpha )m}=0;
\end{equation*}%
\textit{\ }

\textbf{3}. $\bar{\varepsilon}^{\alpha \beta }$ \textit{is a {constant}
non-degenerate symmetric matrix such that}%
\begin{equation}
\bar{\varepsilon}^{\alpha \beta }=\sum \bar{H}_{m}^{(\alpha )}\bar{H}%
_{m}^{(\beta )}\text{, \ \ \ \ \ }\bar{H}_{i}^{(\alpha )}=\bar{\varepsilon}%
^{\alpha \beta }\bar{H}_{(\beta )i}.  \label{eps}
\end{equation}

\textbf{Proof}: can be obtained by the direct verification%
\begin{equation*}
\hat{K}^{is}\hat{M}_{sj}=\delta _{j}^{i}\text{, \ \ \ \ \ \ \ \ \ }\hat{M}%
_{is}\hat{K}^{sj}=\delta _{i}^{j}.
\end{equation*}

\textbf{Corollary}: The Lagrangian%
\begin{equation}
S=\int \left[ \frac{1}{2}\sum \bar{H}_{k}^{2}(\mathbf{r})\frac{r_{\tau }^{k}%
}{r_{x}^{k}}-h(\mathbf{r})\right] dxdt  \label{lagr}
\end{equation}%
determines an \textit{integrable} hydrodynamic type system (\textbf{\ref{sec}%
}), where (cf. (\textbf{\ref{inv}}))%
\begin{equation}
H_{i}=\bar{H}_{i}^{(\beta )}q_{\beta }\text{, \ \ \ \ \ \ \ }q_{\beta }=\sum 
\bar{H}_{(\beta )m}H_{m}\text{, \ \ \ \ \ \ \ }\partial _{i}q_{\beta }=\psi
_{i}\bar{H}_{(\beta )i}.  \label{ina}
\end{equation}%
It means that the Hamiltonian density $h$ (see (\textbf{\ref{zak}})) is
determined by the linear system%
\begin{equation}
\partial _{i}\psi _{k}=\beta _{ki}\psi _{i}\text{,\ \ \ }i\neq k\text{, \ \
\ \ \ \ }\psi _{i}=\partial _{i}H_{i}+\underset{m\neq i}{\sum }\beta
_{im}H_{m},  \label{sup}
\end{equation}%
where $\partial _{i}h=\psi _{i}\bar{H}_{i}$; the momentum density (cf. (%
\textbf{\ref{mom}}))%
\begin{equation}
\bar{h}=\frac{1}{2}\sum \bar{H}_{m}^{2}  \label{mon}
\end{equation}%
can be obtained by the replacement $\psi _{i}\rightarrow \bar{\psi}_{i}$, $%
H_{i}\rightarrow \bar{H}_{i}$ in the above system, where $\partial _{i}\bar{h%
}=\bar{\psi}_{i}\bar{H}_{i}$. If $\tau \rightarrow t$, $\psi _{i}\rightarrow 
\tilde{\psi}_{i}$, $H_{i}\rightarrow \tilde{H}_{i}$ in the above system (cf.
(\textbf{\ref{odin}})), the above Lagrangian determines the hydrodynamic
type system (\textbf{\ref{first}}), where $\partial _{i}\tilde{h}=\tilde{\psi%
}_{i}\bar{H}_{i}$.

\textbf{Remark}: Since the Lagrangian density 
\begin{equation*}
L=L(\mathbf{r},\mathbf{r}_{x},\mathbf{r}_{t})
\end{equation*}%
determines an energy-momentum tensor, whose components are conservation laws
of the energy and the momentum, respectively%
\begin{equation*}
\partial _{t}\left( L-r_{t}^{k}\frac{\partial L}{\partial r_{t}^{k}}\right)
=\partial _{x}\left( r_{t}^{k}\frac{\partial L}{\partial r_{x}^{k}}\right) 
\text{, \ \ \ \ \ \ \ \ }\partial _{t}\left( r_{x}^{k}\frac{\partial L}{%
\partial r_{t}^{k}}\right) =\partial _{x}\left( L-r_{x}^{k}\frac{\partial L}{%
\partial r_{x}^{k}}\right) ,
\end{equation*}%
then the hydrodynamic type system (\textbf{\ref{sec}}) determined by the
Lagrangian (\textbf{\ref{lagr}}) possesses these conservation laws too,
which are given by (see (\textbf{\ref{mon}}))%
\begin{equation*}
\partial _{\tau }h=\frac{1}{2}\partial _{x}\left( \sum H_{m}^{2}\right) 
\text{, \ \ \ \ \ \ \ \ }\frac{1}{2}\partial _{\tau }\left( \sum \bar{H}%
_{m}^{2}\right) =\partial _{x}\left( \sum \bar{H}_{m}H_{m}-h\right) .
\end{equation*}%
\qquad

The first example of integrable hydrodynamic type systems associated with
the above Lagrangian representation was found in \textbf{\cite{Lagrang}}.
Indeed, under the point transformation (from physical variables to the
Riemann invariants)%
\begin{equation*}
dr^{\pm }=du\pm \sqrt{f^{\prime \prime \prime }(\rho )}d\rho
\end{equation*}%
the Lagrangian (\textbf{\ref{lag}}) reduces to (\textbf{\ref{lagr}}), where
the Lame coefficients are given by%
\begin{equation*}
\bar{H}_{\pm }^{2}=\pm \frac{1}{2\sqrt{f^{\prime \prime \prime }(\rho )}},
\end{equation*}%
and the ideal gas dynamics (\textbf{\ref{ideal}}) reduces to the diagonal
form%
\begin{equation*}
r_{t}^{\pm }=\left( \frac{r^{+}+r^{-}}{2}\pm \varphi (r^{+}-r^{-})\right)
r_{x}^{\pm },
\end{equation*}%
where the function $\varphi (z)$ is given in the implicit form%
\begin{equation*}
\varphi (z)=\rho \sqrt{f^{\prime \prime \prime }(\rho )}\text{, \ \ \ \ \ \
\ }dz=2\sqrt{f^{\prime \prime \prime }(\rho )}d\rho .
\end{equation*}

Taking $N$ \textit{arbitrary} solutions $\psi _{i}^{(\beta )}$ of the
adjoint linear problem (\textbf{\ref{lin}}), the hydrodynamic type system (%
\textbf{\ref{first}}) can be re-written in the Hamiltonian form (see (%
\textbf{\ref{nham}}) and cf. (\textbf{\ref{dn}}))%
\begin{equation}
a_{t}^{\alpha }=\bar{\varepsilon}^{\beta \delta }(w_{\beta }^{\alpha
})_{x}D_{x}^{-1}(w_{\delta }^{\gamma })_{x}\frac{\partial h}{\partial
a^{\gamma }},  \label{cons}
\end{equation}%
where conservation law densities and corresponding fluxes of structural
flows (\textbf{\ref{struc}})%
\begin{equation}
a_{t^{\gamma }}^{\alpha }=(w_{\gamma }^{\alpha }(\mathbf{a}))_{x}
\label{struk}
\end{equation}%
are determined by their derivatives%
\begin{equation*}
\partial _{i}a^{\beta }=\psi _{i}^{(\beta )}\bar{H}_{i}\text{, \ \ \ \ \ \ \
\ \ \ }\partial _{i}w_{\gamma }^{\alpha }=\psi _{i}^{(\alpha )}\bar{H}%
_{(\gamma )i}.
\end{equation*}

\section{\textit{Mixed} bi-Hamiltonian structure}

Let us consider an arbitrary semi-Hamiltonian hydrodynamic type system (%
\textbf{\ref{rim}}). It means, that characteristic velocities $v^{i}(\mathbf{%
r})$ satisfy the integrability condition (see \textbf{\cite{Tsar}})%
\begin{equation*}
\partial _{j}\frac{\partial _{k}v^{i}}{v^{k}-v^{i}}=\partial _{k}\frac{%
\partial _{j}v^{i}}{v^{j}-v^{i}}\text{, \ \ \ \ \ \ \ }i\neq j\neq k.
\end{equation*}%
Following S. Tsarev (see \textbf{\cite{Tsar}}) the Lame coefficients $\bar{H}%
_{k}$ are given by%
\begin{equation}
\partial _{k}\ln \bar{H}_{i}=\frac{\partial _{k}v^{i}}{v^{k}-v^{i}}\text{, \
\ \ \ \ \ \ }i\neq k.  \label{lame}
\end{equation}%
Following G. Darboux (see \textbf{\cite{bibTs:dar1}}) rotation coefficients $%
\beta _{ik}$ are given by (\textbf{\ref{rot}}). Suppose the rotation
coefficients $\beta _{ik}$ associated with a given hydrodynamic type system
satisfy the flatness (\textbf{\ref{flat}}) and anti-flatness (\textbf{\ref%
{antiflat}}) conditions, simultaneously. It means that this hydrodynamic
type system has two Hamiltonian structures%
\begin{equation*}
r_{t}^{i}=\{r^{i},\mathbf{H}_{1}\}_{1}=\{r^{i},\mathbf{H}_{2}\}_{2},
\end{equation*}%
where components of the local (Dubrovin--Novikov) Poisson structure%
\begin{equation*}
\{r^{i},r^{j}\}_{1}=\hat{A}^{ij}\delta (x-x^{\prime })
\end{equation*}%
are given by (\textbf{\ref{lp}})%
\begin{equation*}
\hat{A}^{ii}=\frac{1}{\bar{H}_{i}}D_{x}\frac{1}{\bar{H}_{i}}\text{, \ \ \ \ }%
\hat{A}^{ik}|_{i\neq k}=\frac{1}{\bar{H}_{i}\bar{H}_{k}}(\beta
_{ki}r_{x}^{i}-\beta _{ik}r_{x}^{k})
\end{equation*}%
and components of the local symplectic structure%
\begin{equation*}
\hat{M}_{jk}\{r^{k},r^{i}\}_{2}=\delta _{j}^{i}\delta (x-x^{\prime })
\end{equation*}%
are given by (\textbf{\ref{lsym}})%
\begin{equation*}
\hat{M}_{ii}=\frac{\bar{H}_{i}}{r_{x}^{i}}D_{x}\frac{\bar{H}_{i}}{r_{x}^{i}}%
\text{, \ \ \ \ }\hat{M}_{ik}|_{i\neq k}=\bar{H}_{i}\bar{H}_{k}\left( \frac{%
\beta _{ik}}{r_{x}^{k}}-\frac{\beta _{ki}}{r_{x}^{i}}\right) .
\end{equation*}%
Thus, the recursion operator of the \textit{second} order is a product of
local Hamiltonian and symplectic operators of the \textit{first} orders%
\begin{equation*}
\hat{R}_{k}^{i}=\hat{A}^{ij}\hat{M}_{jk},
\end{equation*}%
and the above hydrodynamic type system has {infinitely many local} \textit{%
Hamiltonian} structures of \textit{all odd} orders%
\begin{equation}
r_{t}^{i}=\hat{A}^{ij}\frac{\delta \mathbf{H}_{1}}{\delta r^{j}}=\hat{A}^{ij}%
\hat{M}_{jk}\hat{A}^{ks}\frac{\delta \mathbf{H}_{0}}{\delta r^{s}}=\hat{A}%
^{ij}\hat{M}_{jk}\hat{A}^{ks}\hat{M}_{sn}\hat{A}^{nm}\frac{\delta \mathbf{H}%
_{-1}}{\delta r^{m}}=...  \label{inf}
\end{equation}%
and {infinitely many local} \textit{symplectic} structures of \textit{all odd%
} orders%
\begin{equation*}
\hat{M}_{ij}r_{t}^{j}=\frac{\delta \mathbf{H}_{2}}{\delta r^{i}}\text{, \ \
\ \ }\hat{M}_{ij}\hat{A}^{jk}\hat{M}_{kn}r_{t}^{n}=\frac{\delta \mathbf{H}%
_{3}}{\delta r^{i}}\text{, \ \ \ \ \ }\hat{M}_{ij}\hat{A}^{jk}\hat{M}_{kn}%
\hat{A}^{ns}\hat{M}_{sm}r_{t}^{m}=\frac{\delta \mathbf{H}_{4}}{\delta r^{i}}%
,...
\end{equation*}

\textbf{Theorem 2}: \textit{The nonlinear PDE system}%
\begin{eqnarray}
\partial _{i}\beta _{jk} &=&\beta _{ji}\beta _{ik}\text{, \ \ \ \ \ \ \ }%
i\neq j\neq k\text{,}  \notag \\
&&  \label{l} \\
\partial _{i}\beta _{ik}+\partial _{k}\beta _{ki}+\underset{m\neq i}{\sum }%
\beta _{mi}\beta _{mk} &=&0\text{, \ \ \ \ }\partial _{i}\beta
_{ki}+\partial _{k}\beta _{ik}+\underset{m\neq i}{\sum }\beta _{im}\beta
_{km}=0\text{, \ \ \ \ }i\neq k  \notag
\end{eqnarray}%
\textit{is an integrable system, which is a consequence of compatibility
conditions following from the over-determined linear PDE system}%
\begin{eqnarray*}
\partial _{i}H_{k} &=&\beta _{ik}H_{i}\text{,\ \ \ \ \ \ }i\neq k\text{, \ \
\ \ \ \ }\lambda H_{i}=\partial _{i}\left( \partial _{i}H_{i}+\underset{%
n\neq i}{\sum }\beta _{in}H_{n}\right) +\underset{m\neq i}{\sum }\beta
_{mi}\left( \partial _{m}H_{m}+\underset{n\neq m}{\sum }\beta
_{mn}H_{n}\right) \text{,} \\
&& \\
\partial _{k}\psi _{i} &=&\beta _{ik}\psi _{k}\text{,\ \ \ \ \ \ }i\neq k%
\text{, \ \ \ \ \ \ }\lambda \psi _{i}=\partial _{i}\left( \partial _{i}\psi
_{i}+\underset{n\neq i}{\sum }\beta _{ni}\psi _{n}\right) +\underset{m\neq i}%
{\sum }\beta _{im}\left( \partial _{m}\psi _{m}+\underset{n\neq m}{\sum }%
\beta _{nm}\psi _{n}\right) .
\end{eqnarray*}

\textbf{Proof}: follows from the above construction.

Indeed, suppose some particular solution $H_{i}^{(0)}$ of the first linear
system (\textbf{\ref{lin}}) is given. Then corresponding solution $\psi
_{i}^{(0)}$ of the adjoint linear system (\textbf{\ref{lin}}) can be
obtained from (\textbf{\ref{sup}})%
\begin{equation*}
\psi _{i}^{(0)}=\partial _{i}H_{i}^{(0)}+\underset{n\neq i}{\sum }\beta
_{in}H_{n}^{(0)}.
\end{equation*}%
The new particular solution $H_{i}^{(1)}$ is given by (\textbf{\ref{lok}})%
\begin{equation*}
H_{i}^{(1)}=\partial _{i}\psi _{i}^{(0)}+\underset{m\neq i}{\sum }\beta
_{mi}\psi _{m}^{(0)}.
\end{equation*}%
Thus, an iteration of the above substitutions%
\begin{equation}
H_{i}^{(k)}=\partial _{i}\psi _{i}^{(k+1)}+\underset{m\neq i}{\sum }\beta
_{mi}\psi _{m}^{(k+1)}\text{,\ \ }\psi _{i}^{(k+1)}=\partial
_{i}H_{i}^{(k+1)}+\underset{n\neq i}{\sum }\beta _{in}H_{n}^{(k+1)}\text{,\
\ }k=0,1,2,...  \label{j}
\end{equation}%
leads to \textit{semi-infinite} series of solutions of linear problems (%
\textbf{\ref{lin}}). It means, that corresponding commuting flows and
conservation law densities can be found too%
\begin{equation*}
r_{t^{k}}^{i}=\frac{H_{i}^{(k)}}{\bar{H}_{i}}r_{x}^{i}\text{, \ \ \ \ \ \ }%
dh^{(k)}=\sum \psi _{i}^{(k)}\bar{H}_{i}dr^{i}.
\end{equation*}

Since the iterations (\textbf{\ref{j}}) are invertible (see (\textbf{\ref%
{inv}}) and (\textbf{\ref{ina}})), another semi-infinite series can be found
in quadratures%
\begin{equation*}
H_{i}^{(-k-1)}=\bar{H}_{i}^{(\beta )}q_{\beta }^{(-k)}\text{, \ \ \ }%
\partial _{i}q_{\beta }^{(-k)}=\psi _{i}^{(-k)}\bar{H}_{(\beta )i}\text{, \
\ \ }\psi _{i}^{(-k)}=\bar{\psi}_{i}^{(\beta )}\frac{\partial h^{(-k)}}{%
\partial a^{\beta }}\text{, \ \ \ }\partial _{i}\frac{\partial h^{(-k)}}{%
\partial a^{\beta }}=\bar{\psi}_{(\beta )i}H_{i}^{(-k)}.
\end{equation*}

Let structural flows (\textbf{\ref{struk}}) be written via flat coordinates
of the first local Hamiltonian structure of the Dubrovin--Novikov type (see (%
\textbf{\ref{dn}}))%
\begin{equation*}
a_{t^{\gamma }}^{\alpha }=\bar{g}^{\alpha \beta }D_{x}\frac{\partial
h_{\gamma }}{\partial a^{\beta }},
\end{equation*}%
where%
\begin{equation*}
\partial _{i}a^{\beta }=\bar{\psi}_{i}^{(\beta )}\bar{H}_{i}\text{, \ \ \ \
\ }\partial _{i}w_{\gamma }^{\alpha }=\bar{\psi}_{i}^{(\alpha )}\bar{H}%
_{(\gamma )i}\text{, \ \ \ \ \ \ }\bar{H}_{(\gamma )i}=\partial _{i}\psi
_{(\gamma )i}+\underset{m\neq i}{\sum }\beta _{mi}\psi _{(\gamma )m}\text{,
\ \ \ \ }\partial _{i}h_{\gamma }=\psi _{(\gamma )i}\bar{H}_{i}.
\end{equation*}%
Then an integrable hierarchy of hydrodynamic type systems can be written in
the bi-Hamiltonian form%
\begin{equation}
a_{t^{k}}^{\alpha }=\bar{g}^{\alpha \beta }D_{x}\frac{\partial h^{(k)}}{%
\partial a^{\beta }}=\bar{\varepsilon}^{\beta \delta }(w_{\beta }^{\alpha
})_{x}D_{x}^{-1}(w_{\delta }^{\gamma })_{x}\frac{\partial h^{(k+1)}}{%
\partial a^{\gamma }}.  \label{mokh}
\end{equation}

\textbf{Remark}: More general \textit{compatible} bi-Hamiltonian structure
recently was investigated in \textbf{\cite{Mokh}}. However, the nonlocal
Hamiltonian operator (\textbf{\ref{nham}}) can be effectively constructed
for any given flat metric namely in the above (flat--anti-flat) case.
Moreover, just in this \textit{mixed} case a corresponding \textit{recursion
operator is \textbf{local}}.

Indeed, since%
\begin{equation*}
\bar{\varepsilon}_{\eta \zeta }(U^{\eta \beta }\bar{g}_{\beta \gamma
})D_{x}(U^{\zeta \mu }\bar{g}_{\mu \sigma })a_{t^{k}}^{\sigma }=\frac{%
\partial h^{(k+1)}}{\partial a^{\gamma }},
\end{equation*}%
where%
\begin{equation*}
U^{\alpha \gamma }(w_{\gamma \beta })_{x}=\delta _{\beta }^{\alpha }\text{,
\ \ \ \ \ \ }(w_{\gamma \beta })_{x}U^{\beta \alpha }=\delta _{\gamma
}^{\alpha }\text{, \ \ \ \ \ \ }U^{\alpha \gamma }=\sum \frac{\bar{\psi}%
_{m}^{(\gamma )}\bar{H}_{m}^{(\alpha )}}{r_{x}^{m}}\text{,\ \ \ \ \ \ }%
\partial _{i}w_{\gamma \beta }=\bar{\psi}_{(\gamma )i}\bar{H}_{(\beta )i},
\end{equation*}%
then an integrable hierarchy of hydrodynamic type systems possesses the 
\textit{\textbf{local} Hamiltonian structure of the \textbf{third} order}
given by%
\begin{equation}
a_{t^{k}}^{\alpha }=\bar{\varepsilon}_{\eta \zeta }D_{x}U^{\eta \alpha
}D_{x}U^{\zeta \sigma }D_{x}\frac{\partial h^{(k-1)}}{\partial a^{\sigma }}.
\label{nemokh}
\end{equation}

\textbf{Remark}: In $N$ component case, an inverse matrix $U^{\beta \gamma }(%
\mathbf{a,a}_{x})$ is a very complicated expression with respect to the
first order derivatives. By this reason, the Riemann invariants $r^{k}$ (see
(\textbf{\ref{inf}})) are most appropriate coordinates (for instance, in the
Whitham theory for an arbitrary genus higher than $0$), except the case when
the Riemann invariants cannot be found explicitly, for instance for a
dispersionless limit of integrable dispersive systems (see e.g. \textbf{\cite%
{Maks+algebr}}).

\textbf{Remark}: One of very powerful approaches in an integrability of the
nonlinear PDE system (\textbf{\ref{non}}) was presented in \textbf{\cite%
{BogdFer}}. Two linear PDE systems (\textbf{\ref{lin}}) should be related by
the linear ODE system (with respect to each fixed index $i$)%
\begin{equation*}
H_{i}=[c_{i}\partial _{i}^{n}+...+c_{n}(\mathbf{r})]\psi _{i}.
\end{equation*}%
If $n=1$, this is exactly already well-known flat case (\textbf{\ref{lok}}).
The case considered in \textbf{\cite{BogdFer}} can be restricted on a
compatible couple of the above transformations%
\begin{equation*}
H_{i}^{(k+1)}=[c_{(0)i}\partial _{i}^{n}+...+c_{(n)i}(\mathbf{r})]\psi
_{i}^{(k)}\text{, \ \ \ \ \ \ \ \ \ }\psi _{i}^{(k)}=[\tilde{c}%
_{(0)i}\partial _{i}^{n}+...+\tilde{c}_{(n)i}(\mathbf{r})]H_{i}^{(k)}
\end{equation*}%
generalizing the case considered in this Section.

\textbf{Remark}: The above transformation of the \textit{second} order (see (%
\textbf{\ref{l}}) and (\textbf{\ref{j}}))%
\begin{equation*}
H_{i}^{(k+1)}=(\partial _{i}^{2}+...)H_{i}^{(k)}
\end{equation*}%
is a first example of more general transformations%
\begin{equation*}
H_{i}^{(k+1)}=(\partial _{i}^{n}+...)H_{i}^{(k)}
\end{equation*}%
never studied except $n=1$ (see below, Section \textbf{7}).

\textbf{Remark}: Thus, the first ``puzzle'' hidden in \textbf{\cite{NO}} is
a consequence of the existence of a mixed bi-Hamiltonian structure discussed
in this section. It means, that the sub-class of so-called ``separable''
Hamiltonians for two component hydrodynamic type systems introduced in 
\textbf{\cite{NO}} has pure differential-geometric meaning via the language
of conjugate curvilinear coordinate nets in $N$ component case.

\textbf{Remark}: The nonlinear PDE system (\textbf{\ref{l}}) is well-known
in classical differential geometry (see e.g. \textbf{\cite{bibTs:bianchi}}).
See also recent investigation in \textbf{\cite{Terng}}, where (\textbf{\ref%
{l}}) is called $O(2N)/O(N)\times O(N)-$system (see also the \textit{%
generalized wave equation} in \textbf{\cite{Tenenblat}}). This nonlinear PDE
system (\textbf{\ref{l}}) is a natural generalization of the Egorov case
(see below Section \textbf{7}): if rotation coefficients $\beta _{ik}$ are
symmetric, then (\textbf{\ref{l}}) reduces to well-known system describing
Egorov flat metrics.

\section{Multi-Hamiltonian structures}

A comparison nonlinear and linear PDE systems ((\textbf{\ref{non}}), (%
\textbf{\ref{lin}}), (\textbf{\ref{lok}}) and (\textbf{\ref{flat}}))
describing orthogonal curvilinear coordinate nets $\beta _{ik}$ with
corresponding nonlinear and linear PDE systems ((\textbf{\ref{antiflat}})
and (\textbf{\ref{sup}})) describing anti-orthogonal curvilinear coordinate
nets $\tilde{\beta}_{ik}$ leads to the simple identity%
\begin{equation}
\tilde{\beta}_{ik}=\beta _{ki}.  \label{mir}
\end{equation}

\textbf{Definition}: \textit{Corresponding curvilinear coordinate nets are
said to be {\textbf{mirrored}}}.

Since a description of orthogonal curvilinear coordinate nets is \textit{%
equivalent} to a description of anti-orthogonal curvilinear coordinate nets,
then, in fact, all known theorems in the theory of orthogonal curvilinear
coordinate nets can be simply reformulated for anti-orthogonal case.

\textbf{Interpretation}: Any integrable (semi-Hamiltonian) hydrodynamic type
system has an infinite set of conservation laws and commuting flows.

Let us write $N$ arbitrary conservation laws for $N-1$ arbitrary nontrivial
commuting flows in the potential form%
\begin{equation}
d\left( 
\begin{array}{c}
y_{1} \\ 
y_{2} \\ 
... \\ 
y_{N}%
\end{array}%
\right) =\left| 
\begin{array}{cccc}
h_{11} & h_{12} & ... & h_{1N} \\ 
h_{21} & h_{22} & ... & h_{2N} \\ 
. & . & . & . \\ 
h_{N1} & h_{N2} & ... & h_{NN}%
\end{array}%
\right| d\left( 
\begin{array}{c}
t_{1} \\ 
t_{2} \\ 
... \\ 
t_{N}%
\end{array}%
\right) .  \label{zer}
\end{equation}%
It means, that $N-1$ commuting flows%
\begin{equation}
r_{t^{k}}^{i}=\frac{H_{(k)i}}{H_{(1)i}}r_{t^{1}}^{i}\text{, \ \ \ \ \ }%
k=2,3,...,N  \label{raz}
\end{equation}%
possess $N$ conservation laws%
\begin{equation*}
\partial _{t^{k}}h_{i1}=\partial _{t^{1}}h_{ik}\text{, \ \ \ \ \ }%
i=1,2,...,N,
\end{equation*}%
where%
\begin{equation*}
\partial _{j}h_{ik}=\psi _{(i)j}H_{(k)j}.
\end{equation*}

Let us construct another set of $N$ conservation laws and $N-1$ commuting
flows determined by the \textit{transposed} matrix%
\begin{equation}
d\left( 
\begin{array}{c}
z_{1} \\ 
z_{2} \\ 
... \\ 
z_{N}%
\end{array}%
\right) =\left| 
\begin{array}{cccc}
h_{11} & h_{21} & ... & h_{N1} \\ 
h_{12} & h_{22} & ... & h_{N2} \\ 
. & . & . & . \\ 
h_{1N} & h_{2N} & ... & h_{NN}%
\end{array}%
\right| d\left( 
\begin{array}{c}
x_{1} \\ 
x_{2} \\ 
... \\ 
x_{N}%
\end{array}%
\right) .  \label{kal}
\end{equation}%
It means, that $N-1$ commuting flows%
\begin{equation}
r_{x^{k}}^{i}=\frac{\psi _{(k)i}}{\psi _{(1)i}}r_{x^{1}}^{i}\text{, \ \ \ \
\ }k=2,3,...,N  \label{dva}
\end{equation}%
possess $N$ conservation laws%
\begin{equation*}
\partial _{x^{k}}h_{1i}=\partial _{x^{1}}h_{ki}\text{, \ \ \ \ \ }%
i=1,2,...,N.
\end{equation*}%
Corresponding rotation coefficients are given by (see (\textbf{\ref{mir}}))%
\begin{equation*}
\tilde{\beta}_{ik}=\frac{\partial _{i}\tilde{H}_{(1)k}}{\tilde{H}_{(1)i}}%
\equiv \frac{\partial _{i}\psi _{(1)k}}{\psi _{(1)i}}=\beta _{ki},
\end{equation*}%
because the new Lame coefficients $\tilde{H}_{(1)i}\equiv \psi _{(1)i}$ (cf.
(\textbf{\ref{raz}}) and (\textbf{\ref{dva}})).

Such families of hydrodynamic type systems are said to be \textit{mirrored}.
The aforementioned relationship between these hydrodynamic type systems (%
\textbf{\ref{zer}}) and (\textbf{\ref{kal}}) is well known in the projective
differential geometry as the ``duality condition'' (see \textbf{\cite%
{Fer+kongr}} and other references therein).

One of the most interesting questions of the classical differential geometry
which has appeared at studying of semi-Hamiltonian systems of hydrodynamical
type is the description of the surfaces admitting not trivial deformations
with preservation of principal directions and principal curvatures. Then the
number of essential parameters on which such deformations depend, is
actually equal to number of various local Hamiltonian structures of a
corresponding hydrodynamical type system \textbf{\cite{viniti}}. Such local
Hamiltonian structures are determined by differential-geometrical Poisson
brackets of the first order (see \textbf{\cite{Dubr+Nov}}). In the same
paper a bi-Hamiltonian structure of the system of averaged one-phase
solutions of Korteweg de Vries equation has been considered. Later
multi-Hamiltonian structures of systems of hydrodynamical type were studied
in \textbf{\cite{Maks+multi}}, \textbf{\cite{Maks+Tsar}}.

\textbf{Example} \textbf{\cite{Maks+Fer}}: The Lame coefficients (\textbf{%
\ref{lame}}) for the hydrodynamic type system 
\begin{equation}
r_{t}^{i}=\left( \sum r^{m}+2r^{i}\right) r_{x}^{i}\text{, \ \ \ \ \ }%
i=1,2,...,N  \label{ellip}
\end{equation}%
are given by%
\begin{equation*}
\bar{H}_{i}=\mu _{i}^{-1/2}(r^{i})\underset{m\neq i}{\prod }%
(r^{i}-r^{m})^{1/2},
\end{equation*}%
where $\mu _{i}(r^{i})$ are arbitrary functions. The rotation coefficients (%
\textbf{\ref{rot}}) are given by%
\begin{equation*}
\beta _{ik}=\frac{\mu _{i}^{1/2}(r^{i})}{\mu _{k}^{1/2}(r^{i})}\frac{\bar{H}%
_{k}^{0}}{\bar{H}_{i}^{0}}\frac{1/2}{r^{i}-r^{k}},
\end{equation*}%
where we use the notation%
\begin{equation}
\bar{H}_{i}^{0}=\underset{m\neq i}{\prod }(r^{i}-r^{m})^{1/2}.  \label{nul}
\end{equation}%
The substitution these rotation coefficients in the flatness condition (%
\textbf{\ref{flat}}) yields the following functional-differential system
(see \textbf{\cite{Tsar}})%
\begin{equation}
\mu _{i}\partial _{i}\beta _{ik}^{0}+\frac{1}{2}\mu _{i}^{\prime }\beta
_{ik}^{0}+\mu _{k}\partial _{k}\beta _{ki}^{0}+\frac{1}{2}\mu _{k}^{\prime
}\beta _{ki}^{0}+\underset{m\neq i,k}{\sum }\mu _{m}\beta _{mi}^{0}\beta
_{mk}^{0}=0,  \label{flatgen}
\end{equation}%
where we use the notation $\beta _{ik}^{0}=\partial _{i}\bar{H}_{k}^{0}/\bar{%
H}_{i}^{0}$.

\textbf{Lemma} \textbf{\cite{Maks+Fer}}: \textit{The above
functional-differential equation has a polynomial solution parameterized by }%
$N+1$ \textit{arbitrary constants}%
\begin{equation*}
\mu
_{i}=C_{0}(r^{i})^{N}+C_{1}(r^{i})^{N-1}+C_{2}(r^{i})^{N-2}+...+C_{N-2}(r^{i})^{2}+C_{N-1}r^{i}+C_{N}.
\end{equation*}

\textbf{Remark} \textbf{\cite{Maks+multi}}: These Riemann invariants $r^{k}$
are nothing but \textit{elliptic} coordinates (see \textbf{\cite{bibTs:dar1}}%
).

\textbf{Theorem 3} \textbf{\cite{Maks+Fer}}: \textit{Semi-Hamiltonian
hydrodynamic type system cannot possess more than }$N+1$ \textit{local
Hamiltonian structures of the Dubrovin--Novikov type}.

Thus, the above hydrodynamic type system (\textbf{\ref{ellip}}) has a
maximum of local Hamiltonian structures of the Dubrovin--Novikov type.

\textbf{Remark}: Bi-Hamiltonian structures of the Dubrovin--Novikov type are
described in the set of publications (see \textbf{\cite{Dubr}}, \textbf{\cite%
{Fer+comp}}, \textbf{\cite{Mokh+comp}}, \textbf{\cite{Maks+comp}}).
Semi-Hamiltonian hydrodynamic type systems connected with \textit{multi}%
-orthogonal curvilinear coordinate nets, where the number of corresponding
local Hamiltonian structures of the Dubrovin--Novikov type is less than $N+1$
is not described yet.

Since orthogonal and anti-orthogonal cases are \textit{dual} to each other,
then following theorem is valid.

\textbf{Theorem 4}: \textit{Semi-Hamiltonian hydrodynamic type system cannot
possess more than }$N+1$ \textit{anti-flat nonlocal Hamiltonian structures
(see} (\textbf{\ref{nham}}), (\textbf{\ref{antiflat}})).

\textbf{Example}: The hydrodynamic type system%
\begin{equation}
r_{t}^{i}=\left( \sum r^{m}-2r^{i}\right) r_{x}^{i}\text{, \ \ \ \ \ }%
i=1,2,...,N  \label{antiellip}
\end{equation}%
has\textit{\ }$N+1$ local Lagrangian representations (see (\textbf{\ref{lagr}%
}))%
\begin{equation*}
S_{n}=\int \left[ \frac{1}{2}\sum \frac{(r^{k})^{n-1}}{\underset{m\neq k}{%
\prod }(r^{k}-r^{m})}\frac{r_{t}^{k}}{r_{x}^{k}}-h_{n}(\mathbf{r})\right]
dxdt\text{, \ \ \ \ \ \ \ }n=1,2,...,N+1.
\end{equation*}%
Indeed, the Lame coefficients are given by (cf. (\textbf{\ref{nul}}))%
\begin{equation*}
\bar{H}_{i}^{0}=\underset{m\neq i}{\prod }(r^{i}-r^{m})^{-1/2}.
\end{equation*}%
It means, that corresponding rotation coefficients $\tilde{\beta}_{ik}$ are
mirrored (see (\textbf{\ref{mir}})) to the rotation coefficients $\beta
_{ik} $ associated with elliptic coordinates (see (\textbf{\ref{ellip}})).

\section{The Egorov flat hydrodynamic type systems}

Let us consider the semi-Hamiltonian hydrodynamic type system (\textbf{\ref%
{rim}}) possessing the local Hamiltonian structure of the Dubrovin--Novikov
type (\textbf{\ref{dn}}) and the couple of conservation laws%
\begin{equation}
a_{t}=b_{x}\text{, \ \ \ \ \ \ \ \ }b_{t}=c_{x}.  \label{egor}
\end{equation}

\textbf{Definition}: \textit{The above Hamiltonian hydrodynamic type system
is said to be Egorov flat}.

\textbf{Theorem 5}: \textit{The Egorov flat hydrodynamic type system has {%
\textbf{infinitely many local} Hamiltonian structures of all \textbf{odd}
orders}.}

\textbf{Proof}: If the semi-Hamiltonian hydrodynamic type system (\textbf{%
\ref{rim}}) has the couple of conservation laws (\textbf{\ref{egor}}), then
(see \textbf{\cite{Maks+Tsar}}) $a$ is a potential of the Egorov metric, i.e.%
\begin{equation*}
\partial _{i}a=\bar{H}_{i}^{2}\text{, \ \ \ \ \ \ }\partial _{i}b=\tilde{H}%
_{i}\bar{H}_{i}\text{, \ \ \ \ \ \ \ }\partial _{i}c=\tilde{H}_{i}^{2}
\end{equation*}%
and rotation coefficients $\beta _{ik}$ are symmetric (see \textbf{\cite%
{Egorov}}). It means, that both linear PDE systems (\textbf{\ref{lin}}) 
\textit{coincide} to each other (thus, we can identify $\psi _{k}$ and $%
H_{k} $ below). Also the flatness (\textbf{\ref{flat}}) and anti-flatness (%
\textbf{\ref{antiflat}}) conditions \textit{coincide} and reduce to%
\begin{equation}
\delta \beta _{ik}=0,  \label{diff}
\end{equation}%
where $\delta =\sum \partial /\partial r^{m}$ is a shift operator. Thus, if
the Egorov hydrodynamic type system has a local Hamiltonian structure of the
Dubrovin--Novikov type, then another nonlocal Hamiltonian structure
associated with a local Lagrangian representation exists automatically; and
vice versa.

The \textbf{main statement} of this section: \textit{If rotation coefficients%
} $\beta _{ik}$ \textit{are \textbf{symmetric} and depend on \textbf{%
differences} of Riemann invariants (see} (\textbf{\ref{diff}})) \textit{%
only, then corresponding hydrodynamic type system} (\textbf{\ref{rim}}) 
\textit{has an \textbf{infinite} set of \textbf{local} Hamiltonian
structures of \textbf{all odd} orders}.

\textbf{Remark}: In general case, an infinite set of Hamiltonian structures
can be constructed starting with a couple of given Hamiltonian structures
(one of them must be invertible) due to Magri's theorem (see \textbf{\cite%
{Magri}}). In the Egorov case, we need just one local Hamiltonian structure
of the Dubrovin--Novikov type or one local Lagrangian representation (%
\textbf{\ref{lagr}}).

Indeed, the first transformation (\textbf{\ref{lok}}) (associated with the
flatness condition (\textbf{\ref{flat}})) reduces to (see \textbf{\cite{Tsar}%
}, \textbf{\cite{Tsar+geom}})%
\begin{equation}
H_{i}^{(k)}=\delta H_{i}^{(k+1)};  \label{shift}
\end{equation}%
the second transformation (\textbf{\ref{sup}}) (associated with the
anti-flatness condition (\textbf{\ref{antiflat}})) reduces to (cf. (\textbf{%
\ref{j}}))%
\begin{equation*}
H_{i}^{(k-1)}=\delta H_{i}^{(k)}.
\end{equation*}%
Thus, the transformation of the second order (\textbf{\ref{j}}) factorizes
in a product of the above transformations of the first orders. The
well-known theorem (in fact written by L. Bianchi in \textbf{\cite%
{bibTs:bianchi}}) is a consequence of such a factorization:

\textbf{Theorem 6}: \textit{The nonlinear PDE system
(Bianchi--Lame--Darboux--Egorov) system} (\textbf{\ref{non}}), (\textbf{\ref%
{diff}}) \textit{is integrable. The corresponding linear PDE system (which
is equivalent to }$N$ \textit{linear ODE systems for each fixed index }$i$%
\textit{) is given by}%
\begin{equation}
\partial _{i}H_{k}=\beta _{ik}H_{i}\text{, \ \ \ \ }i\neq k\text{, \ \ \ \ \ 
}\lambda H_{i}=\delta H_{i}.  \label{egorlin}
\end{equation}

\textbf{Proof}: can be obtained by a computation of the compatibility
conditions $\partial _{i}(\partial _{j}H_{k})=\partial _{j}(\partial
_{i}H_{k})$ for\ $i\neq j\neq k$ and $\partial _{i}(\partial
_{k}H_{k})=\partial _{k}(\partial _{i}H_{k})$ for\ $i\neq k$.

Let us consider the consistency of the linear problem (\textbf{\ref{lin}})
with the symmetry operator $\hat{S}=\Sigma s_{k}(r^{k})\partial /\partial
r^{k}$. Under the scaling $dR^{k}(r^{k})=dr^{k}/s_{k}(r^{k})$ this symmetry
operator $\hat{S}$ is equivalent to the shift operator $\delta $.

\textbf{Theorem 7}: \textit{The first linear problem} (\textbf{\ref{lin}}) 
\textit{is compatible with the symmetry operator} $\hat{S}$; \textit{i.e.
the compatibility conditions} $\partial _{i}(\partial _{j}H_{k})=\partial
_{j}(\partial _{i}H_{k})$ \textit{for}\ $i\neq j\neq k$ \textit{and} $%
\partial _{i}(\partial _{k}H_{k})=\partial _{k}(\partial _{i}H_{k})$ \textit{%
for}\ $i\neq k$ \textit{are valid, where}%
\begin{equation*}
\lambda H_{i}=(\hat{S}+c_{i}(r^{i}))H_{i},
\end{equation*}%
\textit{if and only if} 
\begin{equation*}
\hat{S}\beta _{ik}=(c^{i}-c^{k}-s_{i}^{\prime })\beta _{ik}.
\end{equation*}%
\textit{The second linear problem} (\textbf{\ref{lin}}) \textit{is
compatible with the symmetry operator} $\hat{S}$; \textit{i.e. the
compatibility conditions} $\partial _{i}(\partial _{j}\psi _{k})=\partial
_{j}(\partial _{i}\psi _{k})$ \textit{for}\ $i\neq j\neq k$ \textit{and} $%
\partial _{i}(\partial _{k}\psi _{k})=\partial _{k}(\partial _{i}\psi _{k})$ 
\textit{for}\ $i\neq k$ \textit{are valid, where}%
\begin{equation*}
\lambda \psi _{i}=(\hat{S}+c_{i}(r^{i}))\psi _{i},
\end{equation*}%
\textit{if and only if} 
\begin{equation*}
\hat{S}\beta _{ik}=(c^{k}-c^{i}-s_{k}^{\prime })\beta _{ik}.
\end{equation*}

\textbf{Proof}: can be obtained by a straightforward computation.

In the symmetric case ($\beta _{ik}=\beta _{ki}$) the above formulas reduce
to%
\begin{equation}
\lambda H_{i}=(\hat{S}+s_{i}^{\prime }(r^{i})/2)H_{i}\text{, \ \ \ \ \ \ \ \ 
}\hat{S}\beta _{ik}+(s_{i}^{\prime }+s_{k}^{\prime })\beta _{ik}/2=0.
\label{egr}
\end{equation}

\textbf{Theorem 8 \cite{Maks+multi}}, \textbf{\cite{Maks+Tsar}}: \textit{The
semi-Hamiltonian Egorov hydrodynamic type system }(\textbf{\ref{rim}}) 
\textit{has a local Hamiltonian structure of the Dubrovin--Novikov type iff
the symmetry restrictions }(\textbf{\ref{egr}}) \textit{are fulfilled (and
vice versa).}

\textbf{Proof}: Indeed, if the Egorov hydrodynamic type system has a local
Hamiltonian structure of the Dubrovin--Novikov type, then the flatness
condition (see (\textbf{\ref{flatgen}}))%
\begin{equation}
s_{i}\partial _{i}\beta _{ik}+\frac{1}{2}s_{i}^{\prime }\beta
_{ik}+s_{k}\partial _{k}\beta _{ki}+\frac{1}{2}s_{k}^{\prime }\beta _{ki}+%
\underset{m\neq i,k}{\sum }s_{m}\beta _{mi}\beta _{mk}=0  \label{digit}
\end{equation}%
reduces to (\textbf{\ref{egr}}).

Thus, if the functional-differential system (\textbf{\ref{egr}}) has $M$
independent solutions ($M<N+2$), then a corresponding Egorov hydrodynamic
type system has also $M$ local Hamiltonian structures of the
Dubrovin--Novikov type as well as $M$ nonlocal Hamiltonian structures
associated with local Lagrangian representations (\textbf{\ref{lagr}}).

\textbf{Examples}: \textit{The ideal gas dynamics} (\textbf{\ref{ideal}}) is
the Egorov hydrodynamic type system. Indeed, the Egorov pair of conservation
laws (\textbf{\ref{egor}}) is given by%
\begin{equation*}
\rho _{t}=(\rho u)_{x}\text{, \ \ \ \ }(\rho u)_{t}=\left( \rho u^{2}+\rho
^{2}f^{\prime \prime }(\rho )-2\rho f^{\prime }(\rho )+2f(\rho )\right) _{x}.
\end{equation*}%
Just one symmetry operator $\delta =\partial /\partial u$ is connected with
sole local Hamiltonian structure of the Dubrovin--Novikov type and with sole
nonlocal Hamiltonian structure associated with a local Lagrangian
representation (\textbf{\ref{lagr}}). Thus, only one infinite series of
local Hamiltonian structures of all odd orders is connected with the ideal
gas dynamics (\textbf{\ref{ideal}}).

\textit{Whitham equations} (see \textbf{\cite{Maks+multi}}). The averaged $N$
phase solution of the Sinh-Gordon equation is the Egorov hydrodynamic type
system possessing \textit{sole} local Hamiltonian structure of the
Dubrovin--Novikov type. The averaged $N$ phase solution of the Korteweg de
Vries equation is the Egorov hydrodynamic type systems possessing \textit{two%
} local Hamiltonian structures of the Dubrovin--Novikov type. The averaged $%
N $ phase solution of the nonlinear Schr\"{o}dinger equation is the Egorov
hydrodynamic type system possessing \textit{three} local Hamiltonian
structures of the Dubrovin--Novikov type.

Thus, the first Egorov hydrodynamic type system possesses sole infinite
series of local Hamiltonian structures of all odd orders; the second Egorov
hydrodynamic type system possesses two infinite series of local Hamiltonian
structures of all odd orders; the third Egorov hydrodynamic type system
possesses three infinite series of local Hamiltonian structures of all odd
orders.

\textit{Exceptional} linear-degenerate hydrodynamic type system%
\begin{equation*}
r_{t}^{i}=\left( \sum \varepsilon _{m}r^{m}-\varepsilon _{i}\sum
r^{m}\right) r_{x}^{i}\text{, \ \ \ \ \ \ \ }i=1,2,...,N.
\end{equation*}%
Since the Lame coefficients $\bar{H}_{i}=1/\Sigma r^{m}$ (see (\textbf{\ref%
{lame}})), the potential of the Egorov metric $a=-1/\Sigma r^{m}$ and the
rotation coefficients $\beta _{ik}=-1/\Sigma r^{m}$, then the
functional-differential system (\textbf{\ref{digit}}) has $N$ independent
solutions. Indeed, the substitution rotation coefficients in (\textbf{\ref%
{egr}}) leads to the following identity%
\begin{equation*}
\sum s_{k}(r^{k})=\frac{1}{2}(s_{i}^{\prime }+s_{k}^{\prime })\sum r^{m}.
\end{equation*}%
This system has the solution $s_{k}(r^{k})\equiv r^{k}$ and $N-1$ parametric
solution $s_{k}=\limfunc{const}$, where $\Sigma s_{k}=0$. Thus, the above
hydrodynamic type system possesses $N$ infinite series of local Hamiltonian
structures of all odd orders. $N$ parametric local Lagrangian representation
is given by%
\begin{equation*}
S=\int \left[ \sum \frac{s_{0}r^{k}+s_{k}}{\sum r^{m}}\frac{r_{t}^{k}}{%
r_{x}^{k}}-h(\mathbf{r})\right] dxdt\text{.}
\end{equation*}

Let us restrict our consideration on $N-1$ commuting Egorov flat
hydrodynamic type systems written via flat coordinates $a^{\beta }$, whose
first conservation law is given by (see \textbf{\cite{Maks+int}})%
\begin{equation*}
\partial _{t^{\gamma }}a_{1}=\partial _{t^{1}}a_{\gamma },
\end{equation*}%
where $a^{\beta }=\bar{g}^{\beta \gamma }a_{\gamma }$ (see (\textbf{\ref{dn}}%
)).

\textbf{Theorem 9 \cite{Dubr}}: \textit{These set of hydrodynamic type
systems}%
\begin{equation}
\partial _{t^{\gamma }}a_{\beta }=\partial _{t^{1}}\frac{\partial ^{2}F}{%
\partial a^{\beta }\partial a^{\gamma }}  \label{wdvv}
\end{equation}%
\textit{is determined by a solution of the WDVV equation}%
\begin{equation*}
\frac{\partial ^{3}F}{\partial a^{\alpha }\partial a^{\beta }\partial
a^{\gamma }}\bar{g}^{\gamma \eta }\frac{\partial ^{3}F}{\partial a^{\eta
}\partial a^{\zeta }\partial a^{\mu }}=\frac{\partial ^{3}F}{\partial
a^{\alpha }\partial a^{\zeta }\partial a^{\gamma }}\bar{g}^{\gamma \eta }%
\frac{\partial ^{3}F}{\partial a^{\eta }\partial a^{\beta }\partial a^{\mu }}%
.
\end{equation*}

\textbf{Corollary}: \textit{Conservation law densities satisfy the linear
PDE system}%
\begin{equation*}
\bar{g}^{\gamma \eta }\frac{\partial ^{2}h}{\partial a^{\beta }\partial
a^{\gamma }}\frac{\partial ^{3}F}{\partial a^{\eta }\partial a^{\zeta
}\partial a^{\mu }}=\bar{g}^{\gamma \eta }\frac{\partial ^{2}h}{\partial
a^{\mu }\partial a^{\gamma }}\frac{\partial ^{3}F}{\partial a^{\eta
}\partial a^{\zeta }\partial a^{\beta }}.
\end{equation*}%
Taking into account (this is a consequence from (\textbf{\ref{wdvv}}) when $%
\beta =1$)%
\begin{equation*}
\bar{g}_{\alpha \beta }=\frac{\partial ^{3}F}{\partial a^{1}\partial
a^{\beta }\partial a^{\gamma }}
\end{equation*}%
the above linear PDE system reduces (for $\beta =1$) to%
\begin{equation*}
\bar{g}^{\gamma \eta }\frac{\partial ^{2}h}{\partial a^{1}\partial a^{\gamma
}}\frac{\partial ^{3}F}{\partial a^{\eta }\partial a^{\zeta }\partial a^{\mu
}}=\frac{\partial ^{2}h}{\partial a^{\mu }\partial a^{\zeta }}.
\end{equation*}%
Since this linear PDE system is consistent with the symmetry (shift)
operator $\delta =\partial /\partial a^{1}$, then this formula reduces to
iterations (in quadratures) of all conservation law densities (see \textbf{%
\cite{Dubr}})%
\begin{equation*}
\frac{\partial ^{2}h_{\beta }^{(k+1)}}{\partial a^{\mu }\partial a^{\zeta }}=%
\bar{g}^{\gamma \eta }\frac{\partial h_{\beta }^{(k)}}{\partial a^{\gamma }}%
\frac{\partial ^{3}F}{\partial a^{\eta }\partial a^{\zeta }\partial a^{\mu }}%
\text{, \ \ \ \ \ \ }k=0,1,2,...\text{, \ \ \ \ \ }p=1,2,...,N,
\end{equation*}%
where first $N$ conservation law densities $h_{\beta }^{(0)}$ are (for
instance) flat coordinates $a^{\beta }$. Thus, the generating function of
conservation law densities $h(\mathbf{a},\lambda )$ is determined by the
linear PDE system%
\begin{equation}
\lambda \frac{\partial ^{2}h}{\partial a^{\mu }\partial a^{\zeta }}=\bar{g}%
^{\gamma \eta }\frac{\partial h}{\partial a^{\gamma }}\frac{\partial ^{3}F}{%
\partial a^{\eta }\partial a^{\zeta }\partial a^{\mu }},  \label{linegor}
\end{equation}%
whose compatibility conditions (see \textbf{\cite{Dubr}}) lead to the
aforementioned WDVV equation.

\textbf{Theorem 10}: \textit{The above linear PDE system is equivalent to }(%
\textbf{\ref{egorlin}}).

\textbf{Proof}: A differentiation (\textbf{\ref{wdvv}}) with respect to
Riemann invariants (\textbf{\ref{raz}}) leads to the identity (see \textbf{%
\cite{Dubr}})%
\begin{equation*}
\frac{\partial ^{3}F}{\partial a^{\alpha }\partial a^{\beta }\partial
a^{\gamma }}=\sum \frac{\bar{H}_{(\alpha )m}\bar{H}_{(\beta )m}\bar{H}%
_{(\gamma )m}}{\bar{H}_{(1)m}}.
\end{equation*}%
Expressing derivatives of\ $h$ with respect to flat coordinates via
derivatives of $h$ with respect to Riemann invariants (see (\textbf{\ref{zak}%
}), (\textbf{\ref{zuk}}) and (\textbf{\ref{invert}}), where we must identify 
$\bar{\psi}_{(\gamma )i}\equiv \bar{H}_{(\gamma )i}$), (\textbf{\ref{linegor}%
}) reduces to (\textbf{\ref{egorlin}}).

In this case the bi-Hamiltonian structure (\textbf{\ref{mokh}}) reduces to
(see \textbf{\cite{Mokh}})%
\begin{equation*}
a_{t^{k}}^{\alpha }=\bar{g}^{\alpha \beta }D_{x}\frac{\partial h^{(k)}}{%
\partial a^{\beta }}=\bar{g}^{\beta \zeta }\bar{g}^{\alpha \eta }\bar{g}%
^{\gamma \mu }\left( \frac{\partial ^{2}F}{\partial a^{\eta }\partial
a^{\beta }}\right) _{x}D_{x}^{-1}\left( \frac{\partial ^{2}F}{\partial
a^{\zeta }\partial a^{\mu }}\right) _{x}\frac{\partial h^{(k+1)}}{\partial
a^{\gamma }},
\end{equation*}%
because the structural flows (\textbf{\ref{struc}}) (see also (\textbf{\ref%
{struk}})) are given by (\textbf{\ref{wdvv}}) and (cf. (\textbf{\ref{metr}})
with (\textbf{\ref{eps}})) $\bar{g}^{\alpha \beta }\equiv \bar{\varepsilon}%
^{\beta \delta }$. A corresponding \textit{\textbf{local} Hamiltonian
structure of the \textbf{third} order} (\textbf{\ref{nemokh}}) reduces to
(cf. (\textbf{\ref{no}}))%
\begin{equation*}
a_{t^{k}}^{\alpha }=\bar{g}_{\beta \gamma }D_{x}U^{\alpha \beta
}D_{x}U^{\gamma \sigma }D_{x}\frac{\partial h^{(k-1)}}{\partial a^{\sigma }},
\end{equation*}%
where%
\begin{equation*}
U^{\alpha \beta }\left( \frac{\partial ^{2}F}{\partial a^{\beta }\partial
a^{\gamma }}\right) _{x}=\delta _{\gamma }^{\alpha }\text{,\ }\left( \frac{%
\partial ^{2}F}{\partial a^{\alpha }\partial a^{\beta }}\right) _{x}U^{\beta
\gamma }=\delta _{\alpha }^{\gamma }\text{,\ \ }U^{\alpha \beta }=\sum \frac{%
\bar{H}_{m}^{(\alpha )}\bar{H}_{m}^{(\beta )}}{r_{x}^{m}}\text{, \ }\partial
_{i}\frac{\partial ^{2}F}{\partial a^{\beta }\partial a^{\gamma }}=\bar{H}%
_{(\beta )i}\bar{H}_{(\gamma )i}.
\end{equation*}

Thus, the above local Hamiltonian structure of the third order is $N$
component generalization of (\textbf{\ref{no}}).

\textbf{Remark}: Ya. Nutku and P. Olver explained the origin of the second
order recursion operator for the ideal gas dynamics (\textbf{\ref{ideal}})
as a product of two Sheftel--Teshukov first order recursion operators in 
\textbf{\cite{NO}}. E.V. Ferapontov has proved (mentioned in \textbf{\cite%
{Tsar}}, see also \textbf{\cite{Fordy+Gurel}}) that the existence of the
Sheftel--Teshukov recursion operator (see \textbf{\cite{Sheftel}}, \textbf{%
\cite{Teshukov}}) is equivalent to the existence of the symmetry operator $%
\delta $ (see (\textbf{\ref{shift}})) in \textit{appropriate} Riemann
invariants. Indeed, the eigenfunction problem (\textbf{\ref{linegor}}) can
be written in the form%
\begin{equation*}
\lambda \frac{\partial h}{\partial a^{\alpha }}=\bar{g}^{\beta \gamma
}D_{x}^{-1}\left( \frac{\partial ^{2}F}{\partial a^{\alpha }\partial
a^{\beta }}\right) _{x}\frac{\partial h}{\partial a^{\gamma }}.
\end{equation*}%
Thus, the recursion operator (translating the gradient $\partial
h^{(k)}/\partial a^{\alpha }$ to the gradient $\partial h^{(k+1)}/\partial
a^{\alpha }$) is given by%
\begin{equation*}
\hat{Q}_{\alpha }^{\beta }=\bar{g}^{\beta \gamma }D_{x}^{-1}\left( \frac{%
\partial ^{2}F}{\partial a^{\alpha }\partial a^{\gamma }}\right) _{x};
\end{equation*}%
the inverse recursion operator%
\begin{equation*}
\hat{G}_{\beta }^{\alpha }=U^{\alpha \gamma }\bar{g}_{\gamma \beta }D_{x}
\end{equation*}%
is connected with the Sheftel--Teshukov recursion operator (translating
corresponding commuting flows) written via flat coordinates (cf. (\textbf{%
\ref{rek}}))%
\begin{equation}
\hat{R}_{\beta }^{\alpha }=\bar{g}^{\alpha \gamma }D_{x}\hat{G}_{\gamma
}^{\sigma }D_{x}^{-1}\bar{g}_{\sigma \beta }=D_{x}(\bar{g}_{\beta \sigma
}U^{\sigma \alpha })  \label{reka}
\end{equation}%
Thus, this observation (made by Ya. Nutku and P. Olver) is valid for $N$
component Egorov Hamiltonian hydrodynamic type systems (cf. (\textbf{\ref%
{nemokh}})).

\textbf{Example}: Let us consider the couple of commuting flows%
\begin{equation*}
\begin{array}{ccccccc}
a_{t}=b_{x}, &  &  &  &  &  & a_{y}=c_{x}, \\ 
&  &  &  &  &  &  \\ 
b_{t}=\partial _{x}(c+z_{bb}), &  &  &  &  &  & b_{y}=\partial _{x}z_{ab},
\\ 
&  &  &  &  &  &  \\ 
c_{t}=\partial _{x}z_{ab}, &  &  &  &  &  & c_{y}=\partial _{x}z_{aa},%
\end{array}%
\end{equation*}%
where the function $z(a,b)$ satisfies the associativity equation (see 
\textbf{\cite{Dubr}})%
\begin{equation*}
z_{aaa}=z_{abb}^{2}-z_{aab}z_{bbb}.
\end{equation*}%
These Egorov hydrodynamic type systems have \textit{only \textbf{one} local}
Hamiltonian structure%
\begin{eqnarray*}
a_{t} &=&D_{x}\frac{\partial h_{1}}{\partial c}\text{, \ \ \ \ \ \ }%
b_{t}=D_{x}\frac{\partial h_{1}}{\partial b}\text{, \ \ \ \ \ \ \ }%
c_{t}=D_{x}\frac{\partial h_{1}}{\partial a}, \\
&& \\
a_{y} &=&D_{x}\frac{\partial h_{2}}{\partial c}\text{, \ \ \ \ \ \ }%
b_{y}=D_{x}\frac{\partial h_{2}}{\partial b}\text{, \ \ \ \ \ \ \ }%
c_{y}=D_{x}\frac{\partial h_{2}}{\partial a}
\end{eqnarray*}%
of the \textit{Dubrovin--Novikov} type in general case, where $%
h_{1}=bc+z_{b} $, $h_{2}=c^{2}/2+z_{a}$.

These commuting flows can be written in the potential form%
\begin{equation}
d\left( 
\begin{array}{c}
\Omega _{x} \\ 
\Omega _{t} \\ 
\Omega _{y}%
\end{array}%
\right) =\left( 
\begin{array}{ccc}
a & b & c \\ 
b & c+z_{bb} & z_{ab} \\ 
c & z_{ab} & z_{aa}%
\end{array}%
\right) d\left( 
\begin{array}{c}
x \\ 
t \\ 
y%
\end{array}%
\right) ,  \label{matr}
\end{equation}%
where%
\begin{equation*}
d\Omega =\Omega _{x}dx+\Omega _{t}dt+\Omega _{y}dy.
\end{equation*}%
Then a new \textit{local} Hamiltonian operator of the \textit{third} order
(cf. (\textbf{\ref{no}}))%
\begin{equation*}
\hat{B}=\hat{R}^{2}\left( 
\begin{array}{ccc}
0 & 0 & D_{x} \\ 
0 & D_{x} & 0 \\ 
D_{x} & 0 & 0%
\end{array}%
\right) ,
\end{equation*}%
where the recursion operator $\hat{R}$ is a purely differential operator of
the \textit{first} order (cf. (\textbf{\ref{rek}}))%
\begin{equation*}
\hat{R}=D_{x}(W_{x})^{-1}
\end{equation*}%
according to (\textbf{\ref{reka}}), and the matrix $W$ is given by (cf. (%
\textbf{\ref{matr}}))%
\begin{equation*}
W=\left( 
\begin{array}{ccc}
c & b & a \\ 
z_{ab} & c+z_{bb} & b \\ 
z_{aa} & z_{ab} & c%
\end{array}%
\right) .
\end{equation*}

\section{\textit{Second} Nutku--Olver's ``puzzle''}

Let us consider a particular case of the ideal gas dynamics (\textbf{\ref%
{ideal}}) determined by the special choice $f(\rho )=\rho ^{\gamma }/\gamma
(\gamma -1)(\gamma -2)$. The polytropic gas dynamics (see \textbf{\cite%
{Nutku}})%
\begin{equation*}
\rho _{t}=(\rho u)_{x}\text{, \ \ \ \ }u_{t}=\left( \frac{u^{2}}{2}+\frac{%
\rho ^{\gamma -1}}{\gamma -1}\right) _{x}
\end{equation*}%
has \textit{\textbf{three local}} Hamiltonian structures of the \textit{%
Dubrovin--Novikov} type 
\begin{equation*}
\left( 
\begin{array}{c}
\rho \\ 
u%
\end{array}%
\right) _{t}=\left( 
\begin{array}{cc}
\hat{A}_{k}^{11} & \hat{A}_{k}^{12} \\ 
\hat{A}_{k}^{21} & \hat{A}_{k}^{22}%
\end{array}%
\right) \left( 
\begin{array}{c}
\frac{\delta \mathbf{H}_{5-k}}{\delta \rho } \\ 
\frac{\delta \mathbf{H}_{5-k}}{\delta u}%
\end{array}%
\right) \text{, \ \ \ \ \ \ }k=1,2,3,
\end{equation*}%
where the Hamiltonians are%
\begin{equation*}
\mathbf{H}_{4}=\int \left[ \frac{\rho u^{2}}{2}+\frac{\rho ^{\gamma }}{%
\gamma (\gamma -1)}\right] dx\text{, \ \ \ \ \ }\mathbf{H}_{3}=\int \rho udx%
\text{, \ \ \ \ \ }\mathbf{H}_{2}=\int \rho dx\text{, \ \ \ \ \ }\mathbf{H}%
_{1}=\int udx\text{.}
\end{equation*}%
Corresponding components $\hat{A}_{k}^{ij}$ of the above local Hamiltonian
operators are given by 
\begin{equation}
\hat{A}_{1}=\left( 
\begin{array}{cc}
0 & D_{x} \\ 
D_{x} & 0%
\end{array}%
\right) ,  \notag
\end{equation}%
\begin{equation*}
\hat{A}_{2}=\left( 
\begin{array}{cc}
\rho \,D_{x}+D_{x}\,\rho & (\gamma -2)\,D_{x}\,u+u\,D_{x} \\ 
D_{x}\,u+(\gamma -2)\,u\,D_{x} & \rho ^{\gamma -2}D_{x}+D_{x}\,\rho ^{\gamma
-2}%
\end{array}%
\right) ,
\end{equation*}%
\begin{equation*}
\hat{A}_{3}=\left( 
\begin{array}{cc}
u\,\rho \,D_{x}+D_{x}\,u\,\rho & 
\begin{array}{c}
D_{x}\left[ \frac{1}{2}(\gamma -2)u^{2}+\frac{1}{\gamma -1}\rho ^{\gamma -1}%
\right] \\ 
+\left[ \frac{1}{2}u^{2}+\frac{1}{\gamma -1}\rho ^{\gamma -1}\right] D_{x}%
\end{array}
\\ 
\begin{array}{c}
D_{x}\left[ \frac{1}{2}u^{2}+\frac{1}{\gamma -1}\rho ^{\gamma -1}\right] \\ 
+\left[ \frac{1}{2}(\gamma -2)u^{2}+\frac{1}{\gamma -1}\rho ^{\gamma -1}%
\right] D_{x}%
\end{array}
& u\,\rho ^{\gamma -2}\,D_{x}+D_{x}\,u\,\rho ^{\gamma -2}%
\end{array}%
\right) .
\end{equation*}%
Then the first recursion operator%
\begin{equation*}
\hat{R}_{1}=\left( 
\begin{array}{cc}
(\gamma -1)\,u+(\gamma -2)u_{x}D_{x}^{-1} & 2\rho +\rho _{x}D_{x}^{-1} \\ 
2\rho ^{\gamma -2}+(\gamma -2)\,\rho ^{\gamma -3}\rho _{x}D_{x}^{-1} & 
\,(\gamma -1)u+u_{x}D_{x}^{-1}%
\end{array}%
\right)
\end{equation*}%
is a ratio of first two local Hamiltonian structures, i.e. $\hat{A}_{2}=\hat{%
R}_{1}\hat{A}_{1}$. However $\hat{A}_{3}\neq \hat{R}_{1}\hat{A}_{2}$. It
means, that%
\begin{equation*}
\hat{R}_{1}\hat{A}_{2}=2(\gamma -1)\left( 
\begin{array}{cc}
2u\rho D_{x}+(u\rho )_{x} & 
\begin{array}{c}
\lbrack \frac{1}{2}(\gamma -1)u^{2}+2\frac{\rho ^{\gamma -1}}{\gamma -1}%
]D_{x} \\ 
+[\frac{1}{2}(\gamma -2)u^{2}+\frac{\rho ^{\gamma -1}}{\gamma -1}]_{x}%
\end{array}
\\ 
\begin{array}{c}
\lbrack \frac{1}{2}(\gamma -1)u^{2}+2\frac{\rho ^{\gamma -1}}{\gamma -1}%
]D_{x} \\ 
+[\frac{1}{2}u^{2}+\frac{\rho ^{\gamma -1}}{\gamma -1}]_{x}%
\end{array}
& 2u\rho ^{\gamma -2}D_{x}+(u\rho ^{\gamma -2})_{x}%
\end{array}%
\right)
\end{equation*}%
\begin{equation*}
-(\gamma -2)\left( 
\begin{array}{cc}
u_{x}D_{x}^{-1}\rho _{x}+\rho _{x}D_{x}^{-1}u_{x} & u_{x}D_{x}^{-1}u_{x}+%
\rho _{x}D_{x}^{-1}\left( \frac{\rho ^{\gamma -2}}{\gamma -2}\right) _{x} \\ 
\left( \frac{\rho ^{\gamma -2}}{\gamma -2}\right) _{x}D_{x}^{-1}\rho
_{x}+u_{x}D_{x}^{-1}u_{x} & \left( \frac{\rho ^{\gamma -2}}{\gamma -2}%
\right) _{x}D_{x}^{-1}u_{x}+u_{x}D_{x}^{-1}\left( \frac{\rho ^{\gamma -2}}{%
\gamma -2}\right) _{x}%
\end{array}%
\right)
\end{equation*}%
is a nonlocal Hamiltonian operator of the \textit{Ferapontov} type (see 
\textbf{\cite{Fer+trans}}). A comparison of the above nonlocal Hamiltonian
operator $\hat{R}_{1}\hat{A}_{2}$ and the third local Hamiltonian operator $%
\hat{A}_{3}$ leads to the purely \textit{nonlocal} Hamiltonian operator (%
\textbf{\ref{nham}})%
\begin{eqnarray*}
2\frac{\gamma -1}{\gamma -2}\hat{A}_{3}-\frac{1}{\gamma -2}\hat{R}_{1}\hat{A}%
_{2} &=&\left( 
\begin{array}{cc}
u & \rho \\ 
\frac{\rho ^{\gamma -2}}{\gamma -2} & u%
\end{array}%
\right) _{x}D_{x}^{-1}\left( 
\begin{array}{cc}
u & \rho \\ 
\frac{\rho ^{\gamma -2}}{\gamma -2} & u%
\end{array}%
\right) _{x}\left( 
\begin{array}{cc}
0 & 1 \\ 
1 & 0%
\end{array}%
\right) \\
&& \\
&=&W_{x}D_{x}^{-1}W_{x}\sigma ,
\end{eqnarray*}%
where the matrix $W$ is given by the particular choice $f(\rho )=\rho
^{\gamma }/\gamma (\gamma -1)(\gamma -2)$ in (\textbf{\ref{you}}).

Taking into account the identity (see (\textbf{\ref{nham}}), (\textbf{\ref%
{mokh}}), (\textbf{\ref{nemokh}}))%
\begin{equation*}
\hat{B}=\hat{A}_{1}\hat{K}^{-1}\hat{A}_{1},
\end{equation*}%
where $\hat{B}$ is a local Hamiltonian operator of the third order (\textbf{%
\ref{no}}) and $\hat{K}$ is aforementioned nonlocal Hamiltonian operator
connected with the local Lagrangian representation (\textbf{\ref{lag}}), we
obtain an \textit{interesting algebraic relationship}%
\begin{equation}
\hat{B}=\hat{A}_{1}\left( 2\frac{\gamma -1}{\gamma -2}\hat{A}_{3}-\frac{1}{%
\gamma -2}\hat{A}_{2}\hat{A}_{1}^{-1}\hat{A}_{2}\right) ^{-1}\hat{A}_{1}
\label{svyaz}
\end{equation}%
for local Hamiltonian operators found by Ya. Nutku and P. Olver in \textbf{%
\cite{NO}}.

The above relationship%
\begin{equation*}
\hat{K}=2\frac{\gamma -1}{\gamma -2}\hat{A}_{3}-\frac{1}{\gamma -2}\hat{A}%
_{2}\hat{A}_{1}^{-1}\hat{A}_{2}
\end{equation*}%
has a simple differential-geometric interpretation (see \textbf{\cite%
{Fer+trans}}). Any semi-Hamiltonian hydrodynamic type system has an infinite
series of nonlocal Hamiltonian structures of the Ferapontov type. These
Hamiltonian structures are associated with surfaces with a flat normal
bundle. A reconstruction of corresponding surfaces with a flat normal bundle
is a very complicated problem (see \textbf{\cite{Bogd+Fer}}) in general. Let 
$g_{ii}=\bar{H}_{i}^{2}$ be a diagonal metric of the first Hamiltonian
structure (i.e. $\hat{A}_{1}$ is written via Riemann invariants $r^{k}$, see
(\textbf{\ref{lp}})), then $g_{ii}^{(2)}=\bar{H}_{i}^{2}/r^{i}$ is a metric
of the second Hamiltonian structure (i.e. $\hat{A}_{2}$), and $g_{ii}^{(3)}=%
\bar{H}_{i}^{2}/(r^{i})^{2}$ is a metric of the third Hamiltonian structure
(i.e. $\hat{A}_{3}$ and $\hat{A}_{2}\hat{A}_{1}^{-1}\hat{A}_{2}$). Thus, the
third metric is associated with two different submanifolds. It means that
corresponding the two-dimensional Riemann space with a flat normal bundle is
embedded into a four dimensional Riemann space in the second (nonlocal) case.

\textbf{Remark}: Ya. Nutku and P. Olver have found one local Hamiltonian
operator of the third order. However, the polytropic gas dynamics has three
local Hamiltonian structures and three local Lagrangian representations (%
\textbf{\ref{lagr}}) too, because this Egorov hydrodynamic type systems
possesses three symmetry operators (the shift operator $\delta =\Sigma
\partial /\partial r^{m}$, the scaling operator $\hat{S}=\Sigma \partial
/\partial r^{m}$, the projective operator $\hat{P}=\Sigma
(r^{m})^{2}\partial /\partial r^{m}$, see (\textbf{\ref{egr}}) and \textbf{%
\cite{Aksenov}}), where each of them is connected with one local Hamiltonian
structure and with one local Lagrangian representation, simultaneously.
Thus, the polytropic gas dynamics has \textbf{three} \textit{local}
Hamiltonian structures of the \textit{third} order. Corresponding flat
coordinates are found in \textbf{\cite{Lagrang}}. Then two other similar
algebraic relationships (see (\textbf{\ref{svyaz}})) can be found.

\section{Generalizations}

Theory of orthogonal curvilinear coordinate nets associated with a flat
metric can be extended on surfaces with a flat normal bundle \textbf{\cite%
{Fer+trans}}. In such a case two linear PDE systems (\textbf{\ref{lin}}) are
connected by the nonlocal transformation of the first order (cf. (\textbf{%
\ref{lok}}))%
\begin{equation*}
H_{i}=\partial _{i}\psi _{i}+\underset{m\neq i}{\sum }\beta _{mi}\psi
_{m}+\varepsilon _{\alpha \beta }H_{i}^{(\alpha )}p^{\beta }\text{, \ \ \ \
\ \ \ }\partial _{i}p^{(\beta )}=\psi _{i}H_{i}^{(\beta )},
\end{equation*}%
where $\varepsilon _{\alpha \beta }$ is a constant symmetric non-degenerate
matrix, $H_{i}^{(\alpha )}$ are particular solutions of (\textbf{\ref{lin}}%
), the flatness condition replaces by the more general Gauss-Codazzi equation%
\begin{equation*}
\partial _{i}\beta _{ik}+\partial _{k}\beta _{ki}+\underset{m\neq i,k}{\sum }%
\beta _{mi}\beta _{mk}=\varepsilon _{\alpha \beta }H_{i}^{(\alpha
)}H_{k}^{(\beta )}.
\end{equation*}

Thus, an obvious generalization of an anti-flatness condition and a
corresponding nonlocal transformation of the first order between two linear
PDE systems (\textbf{\ref{lin}}) can be given by%
\begin{equation}
\psi _{i}=\partial _{i}H_{i}+\underset{m\neq i}{\sum }\beta
_{mi}H_{m}+\varepsilon _{\alpha \beta }\psi _{i}^{(\alpha )}\tilde{p}^{\beta
}\text{, \ \ \ \ \ \ \ }\partial _{i}\tilde{p}^{(\beta )}=\psi _{i}^{(\beta
)}H_{i},  \label{nonlin}
\end{equation}%
where%
\begin{equation*}
\partial _{i}\beta _{ki}+\partial _{k}\beta _{ik}+\underset{m\neq i,k}{\sum }%
\beta _{im}\beta _{km}=\tilde{\varepsilon}_{\alpha \beta }\psi _{i}^{(\alpha
)}\psi _{k}^{(\beta )}.
\end{equation*}

\textbf{Example}: The hydrodynamic type system (\textbf{\ref{antiellip}})
has $N+1$ local symplectic structures $\hat{M}_{ik}^{(n)}$ related by the
recursion operator (cf. \textbf{\cite{Maks+Fer}})%
\begin{equation*}
\hat{R}=\left( 
\begin{array}{cccc}
r^{1} &  &  & 0 \\ 
& r^{2} &  &  \\ 
. & . & . & . \\ 
0 &  &  & r^{N}%
\end{array}%
\right) -\frac{1}{2}\left( 
\begin{array}{cccc}
r_{x}^{1} & r_{x}^{1} & ... & r_{x}^{1} \\ 
r_{x}^{2} & r_{x}^{2} & ... & r_{x}^{2} \\ 
. & . & . & . \\ 
r_{x}^{N} & r_{x}^{N} & ... & r_{x}^{N}%
\end{array}%
\right) D_{x}^{-1}.
\end{equation*}%
Indeed, it is easy to verify that $\hat{M}_{ik}^{(2)}=\hat{M}_{im}^{(1)}\hat{%
R}_{k}^{m}$, $\hat{M}_{ik}^{(3)}=\hat{M}_{im}^{(2)}\hat{R}_{k}^{m}$,..., $%
\hat{M}_{ik}^{(N+1)}=\hat{M}_{im}^{(N)}\hat{R}_{k}^{m}$ are local. However, $%
\hat{M}_{ik}^{(N+2)}=\hat{M}_{im}^{(N+1)}\hat{R}_{k}^{m}$ is no longer
local. Then one can check that corresponding relationships between linear
PDE systems (\textbf{\ref{lin}}) associated with higher symplectic
structures $\hat{M}_{ik}^{(N+1+n)}$ are given by (\textbf{\ref{nonlin}}).

Some nonlocal Hamiltonian structures of the Ferapontov type can be \textit{%
inverted}. However, corresponding symplectic structures are nonlocal too
(see \textbf{\cite{Malt+Nov}}). Thus, a list of all possible local
Hamiltonian structures associated with semi-Hamiltonian hydrodynamic type
systems is \textbf{exhausted} in this paper.

\textbf{Remark}: The paper \textbf{\cite{Fer+kongr}} particularly is devoted
to a connection of Temple's sub-class with a linear-degenerate sub-class of
hydrodynamic type systems. For instance, the linear-degenerate hydrodynamic
type system (see \textbf{\cite{FerMaxNon}})%
\begin{equation*}
r_{t}^{i}=\left( \sum r^{m}-r^{i}\right) r_{x}^{i}
\end{equation*}%
is \textit{mirrored} to Temple's hydrodynamic type system%
\begin{equation*}
r_{y}^{i}=\left( \sum r^{m}+r^{i}\right) r_{z}^{i},
\end{equation*}%
whose commuting flow is well-known electrophoresis hydrodynamic type system
(see \textbf{\cite{Babs}})%
\begin{equation*}
r_{\tau }^{i}=\left( r^{i}\prod r^{m}\right) ^{-1}r_{z}^{i}.
\end{equation*}%
The first hydrodynamic type system has an infinite series of nonlocal
Hamiltonian structures parameterized by $N$ arbitrary functions of a single
variable (see \textbf{\cite{FerMaxNon}}); the second hydrodynamic type
system (as well as the third one) has another infinite series of nonlocal
Hamiltonian structures parameterized by $N$ arbitrary functions of a single
variable (see \textbf{\cite{Fer+trans}}). Thus, the construction presented
in the Section \textbf{5} can be extended on a nonlocal (Ferapontov) case 
\textbf{\cite{Fer+trans}}.

\textbf{Remark}: The \textit{mixed} case (cf. (\textbf{\ref{l}}))%
\begin{eqnarray*}
\partial _{i}\beta _{jk} &=&\beta _{ji}\beta _{ik}\text{, \ \ \ \ \ \ \ }%
i\neq j\neq k, \\
&& \\
\partial _{i}\beta _{ik}+\partial _{k}\beta _{ki}+\underset{m\neq i,k}{\sum }%
\beta _{mi}\beta _{mk} &=&\varepsilon _{\alpha \beta }H_{i}^{(\alpha
)}H_{k}^{(\beta )}\text{, \ \ \ \ \ }\partial _{i}\beta _{ki}+\partial
_{k}\beta _{ik}+\underset{m\neq i,k}{\sum }\beta _{im}\beta _{km}=\tilde{%
\varepsilon}_{\alpha \beta }\psi _{i}^{(\alpha )}\psi _{k}^{(\beta )}
\end{eqnarray*}%
was considered in \textbf{\cite{Tenenblat}} for special simple choice $%
\varepsilon _{\alpha \beta }=\delta _{\alpha \beta }$ and $\tilde{\varepsilon%
}_{\alpha \beta }=0$. Such a case is called the \textit{generalized
Sinh-Gordon equation} (may be better to use another notation: the \textit{%
generalized Bonnet equation}, see a corresponding comment, for instance, in 
\textbf{\cite{Zakh}}) associated with the symmetric spaces $%
SO(2N,1)/SO(N)\times SO(N,1)$ (see details in \textbf{\cite{Tenenblat}}, 
\textbf{\cite{Terng}}). Thus, the above nonlinear system is associated with
an isometric immersion of $N-$dimensional submanifolds into a
pseudo-Euclidean space.

\section{Non-hydrodynamic type systems}

Any semi-Hamiltonian hydrodynamic type system (\textbf{\ref{first}}) has an
infinite set of commuting hydrodynamic type systems (\textbf{\ref{sec}})
parameterized by $N$ arbitrary functions of a single variable (see \textbf{%
\cite{Tsar}}, \textbf{\cite{Tsar+geom}}). However, higher commuting flows%
\begin{equation}
u_{y}^{i}=f_{1}^{i}(u,u_{x},u_{xx})\text{, \ \ \ \ \ \ \ }%
u_{z}^{i}=f_{2}^{i}(u,u_{x},u_{xx},u_{xxx}),...  \label{higher}
\end{equation}%
can be constructed in some cases (see \textbf{\cite{Ganzha}}, \textbf{\cite%
{Sheftel}}, \textbf{\cite{Teshukov}}). If some hydrodynamic type system has
the local Lagrangian representation (\textbf{\ref{lagr}}), then its higher
commuting flows (\textbf{\ref{higher}}) are determined by similar Lagrangians%
\begin{equation*}
S=\int \left[ \frac{1}{2}\sum \bar{H}_{k}^{2}(\mathbf{r})\frac{r_{\tau }^{k}%
}{r_{x}^{k}}-h(\mathbf{r},\mathbf{r}_{x},\mathbf{r}_{xx},\mathbf{r}%
_{xxx},...)\right] dxdt.
\end{equation*}%
Corresponding higher order conservation law densities also were investigated
in \textbf{\cite{Tsar}}, \textbf{\cite{Veroski}}).

\textbf{Theorem 11} \textbf{\cite{Tsar}}: \textit{A semi-Hamiltonian
hydrodynamic type system has the conservation law density}%
\begin{equation*}
h(\mathbf{r},\mathbf{r}_{x})=\sum \frac{\bar{H}_{k}^{2}}{r_{x}^{k}}
\end{equation*}%
\textit{if and only if the extra condition}%
\begin{equation*}
\sum \bar{H}_{k}^{2}\partial _{k}v^{k}=0
\end{equation*}%
\textit{\ is fulfilled}.

Taking into account (\textbf{\ref{first}}), the above constraint reduces to%
\begin{equation*}
\sum (\bar{H}_{k}\partial _{k}\tilde{H}_{k}-\tilde{H}_{k}\partial _{k}\bar{H}%
_{k})=0
\end{equation*}

\textbf{Corollary}: If rotation coefficients $\beta _{ik}$ are symmetric and
depend on differences of Riemann invariants (see (\textbf{\ref{diff}})) only
(the Egorov flat case), then the above constraint reduces to%
\begin{equation*}
\sum \bar{H}_{k}\tilde{\psi}_{k}=\sum \tilde{H}_{k}\bar{\psi}_{k},
\end{equation*}%
where (see (\textbf{\ref{sup}})) $\tilde{\psi}_{k}=\delta \tilde{H}_{k}$ and 
$\bar{\psi}_{k}=\delta \bar{H}_{k}$.

\textbf{Lemma}: \textit{The set of commuting hydrodynamic type systems} (%
\textbf{\ref{wdvv}}) \textit{possesses an infinite series of \textbf{higher}
conservation laws and \textbf{higher} commuting flows}.

\textbf{Proof}: The Lame coefficients $H_{(k)i}$ of the Egorov flat
hydrodynamic type systems (\textbf{\ref{wdvv}}) (written in Riemann
invariants (\textbf{\ref{raz}})) depend on differences of Riemann
invariants. Thus, the above constraint is fulfilled.

This infinite series can be constructed iteratively (see (\textbf{\ref{inf}}%
)). The first commuting flow of the \textbf{third} order is given by the
above Hamiltonian $\mathbf{H}_{1}=\int h(\mathbf{r},\mathbf{r}_{x})dx$%
\begin{equation*}
r_{t}^{i}=\hat{A}^{ij}\frac{\delta \mathbf{H}_{1}}{\delta r^{j}}.
\end{equation*}%
All \textit{higher} commuting flows and \textit{higher} conservation law
densities can be obtained due to the second Hamiltonian structure (see (%
\textbf{\ref{inf}}))%
\begin{equation*}
\frac{\delta \mathbf{H}_{2}}{\delta r^{i}}=\hat{M}_{ij}r_{t}^{j}\text{, \ \
\ \ }\frac{\delta \mathbf{H}_{3}}{\delta r^{i}}=\hat{M}_{ij}\hat{A}^{jk}\hat{%
M}_{kn}r_{t}^{n}\text{, \ \ \ \ \ }\frac{\delta \mathbf{H}_{4}}{\delta r^{i}}%
=\hat{M}_{ij}\hat{A}^{jk}\hat{M}_{kn}\hat{A}^{ns}\hat{M}_{sm}r_{t}^{m},...
\end{equation*}

Thus, this result also generalizes the corresponding construction presented
in \textbf{\cite{NO}}.

\textbf{Remark}: All these higher commuting flows (see (\textbf{\ref{higher}}%
)) are \textit{integrable} ``dispersive'' systems with a \textit{rational}
dependence on first derivatives. They are integrable, because they possess
infinitely many Hamiltonian structures, conservation laws and commuting
flows. Their solutions should be considered elsewhere.

\section{Conclusion and outlook}

In this paper we were able to find answers on a set of important questions.

\begin{itemize}
\item \textbf{Lagrangian formulation for nonlocal Hamiltonian structures of
hydrodynamic type systems associated with the anti-flatness condition is
presented.}

\item \textbf{Existence of two Hamiltonian structures associated with
flatness and anti-flatness conditions implies an infinite set of local
Hamiltonian structures of odd higher orders.}

\item \textbf{Description of multi-Hamiltonian structures associated with
anti-flatness conditions is equivalent to the description of local
multi-Hamiltonian structures of the Dubrovin--Novikov type.}

\item \textbf{Existence of one local Hamiltonian structure for the Egorov
hydrodynamic type system implies the existence of infinitely many local
Hamiltonian structures of odd higher degrees.}

\item \textbf{Non-local Hamiltonian structures associated with the
anti-flatness condition are extended on an arbitrary ``co-dimension''.}
\end{itemize}

The problem of description of local Hamiltonian structures of odd orders was
formulated by S.P. Tsarev (see the end of Dr. of Science Thesis). Below we
formulate a most general conjecture, which should be investigated elsewhere.

The theory of local differential-geometric Poisson brackets%
\begin{equation}
\{u^{i}(x),u^{j}(x^{\prime })\}=A^{ij}(\mathbf{u},\mathbf{u}_{x},\mathbf{u}%
_{xx},\mathbf{...})\delta (x-x^{\prime }),  \label{gena}
\end{equation}%
where%
\begin{equation}
A^{ij}(\mathbf{u},\mathbf{u}_{x},\mathbf{u}_{xx},\mathbf{...})=g^{ij}(%
\mathbf{u})D_{x}^{n}+b_{k}^{ij}(\mathbf{u})u_{x}^{k}D_{x}^{n-1}+(c_{k}^{ij}(%
\mathbf{u})u_{xx}^{k}+c_{km}^{ij}u_{x}^{k}u_{x}^{m})D_{x}^{n-2}+...
\label{!}
\end{equation}%
starts from the first order (see \textbf{\cite{Dubr+Nov}})%
\begin{equation*}
\{u^{i}(x),u^{j}(x^{\prime })\}=\left( g^{ij}(\mathbf{u})D_{x}+b_{k}^{ij}(%
\mathbf{u})u_{x}^{k}\right) \delta (x-x^{\prime })
\end{equation*}%
connected with hydrodynamic type systems%
\begin{equation*}
u_{t}^{i}=v_{k}^{i}(\mathbf{u})u_{x}^{k}\text{, \ \ \ \ \ \ \ \ }k=1,2,...,N.
\end{equation*}%
Later these Poisson brackets were generalized on nonlocal case (see \textbf{%
\cite{Fer+trans}}, \textbf{\cite{Malt+Nov}})%
\begin{equation}
\{u^{i}(x),u^{j}(x^{\prime })\}=\left( A^{ij}(\mathbf{u},\mathbf{u}_{x},%
\mathbf{u}_{xx},\mathbf{...)+}\sum \varepsilon ^{\alpha \beta }w_{\alpha
}^{i}D_{x}^{-1}w_{\beta }^{j}\right) \delta (x-x^{\prime }),  \label{long}
\end{equation}%
where functions $u^{i}(x,t)$ simultaneously satisfy commuting flows%
\begin{equation*}
u_{t^{\alpha }}^{i}=w_{\alpha }^{i}(\mathbf{u},\mathbf{u}_{x},\mathbf{u}%
_{xx},\mathbf{...).}
\end{equation*}%
Moreover, in general case the existence of such nonlocal Hamiltonian
structure means integrability of corresponding PDE system.

Just the Poisson brackets of the first order (see \textbf{\cite{Fer+trans}})%
\begin{equation}
\{u^{i}(x),u^{j}(x^{\prime })\}=\left( g^{ij}(\mathbf{u})D_{x}+b_{k}^{ij}(%
\mathbf{u})u_{x}^{k}\mathbf{+}\sum \varepsilon ^{\alpha \beta }w_{(\alpha
)k}^{i}u_{x}^{k}D_{x}^{-1}w_{(\beta )m}^{j}u_{x}^{m}\right) \delta
(x-x^{\prime })  \label{f}
\end{equation}%
are connected with hydrodynamic type systems.

This paper particularly deals with the local Poisson brackets (\textbf{\ref%
{gena}}), where the operator $A^{ij}(\mathbf{u},\mathbf{u}_{x},\mathbf{u}%
_{xx},\mathbf{...)}$ is given by (\textbf{\ref{inf}}) (cf. (\textbf{\ref{!}}%
)).

The \textbf{main conjecture} of this paper is following: a most general
Poisson bracket associated with an integrable hydrodynamic type system is
given by (\textbf{\ref{long}}), where the local part (cf. (\textbf{\ref{!}}%
)) $A^{ij}(\mathbf{u},\mathbf{u}_{x},\mathbf{u}_{xx},\mathbf{...)}$ is
determined by similar \textit{quasi-rational} functions as in (\textbf{\ref%
{inf}}); the nonlocal part is exactly as in (\textbf{\ref{f}}).

\textbf{Remark}: Aforementioned differential-geometric Poisson brackets are
invariant under an arbitrary point transformation $\tilde{u}^{i}=\tilde{u}%
^{i}(\mathbf{u})$, while the new Poisson brackets (\textbf{\ref{inf}}) are
not invariant. \textit{Quasi-rational} function means \textit{rational} with
respect to \textit{higher} derivatives.

\section*{Acknowledgement}

I am indebted to Eugeni Ferapontov for his best suggestion to consider the
Lagrangian representation (\textbf{\ref{lag}}) for the ideal gas dynamics (%
\textbf{\ref{ideal}}) via Riemann invariants.

I am especially grateful to Sergey Tsarev suggested to use the label
``anti-flatness'' for the condition (\textbf{\ref{antiflat}}). Also I would
like to thank S.P. Novikov for stimulating discussions.%


\addcontentsline{toc}{section}{References}

\begin{thebibliography}{99}
\bibitem{Fer+kongr} \emph{S.I. Agafonov, E.V. Ferapontov}, \newblock Theory
of congruences and systems of conservation laws, J. of Math. Sci., \textbf{94%
} No. 5 (1999) 1748--1792. \emph{S.I. Agafonov, E.V. Ferapontov}, \newblock %
Systems of conservation laws from the point of view of projective theory of
congruences, Izvestia RAN (Math.), \textbf{60} (1996) 1--30.

\bibitem{Aksenov} \emph{A.V. Aksenov}, \newblock Symmetries and relations
between solutions of a class of Euler-Poisson-Darboux equations. (Russian)
Dokl. Akad. Nauk. (Reports of RAS), \textbf{381} No. 2 (2001) 176--179.

\bibitem{bibTs:Bao} \emph{D.~Bao, J.~Marsden, R.~Walton}, \newblock The
Hamiltonian structure of general relativistic perfect fluid, Comm. Math.
Phys., \textbf{99} (1985) 319--345.

\bibitem{Babs} \emph{V.G.~Babskii, M.Yu.~Zhukov, V.I.~Yudovich}, \newblock %
Mathematical theory of electrophoresis. Kiev, 1983; Plenum Press, N.Y., 1988.

\bibitem{bibTs:bianchi} \emph{L. Bianchi}, \newblock Sisteme tripli
ortogonali, Ed. Cremonese, Roma (1955).

\bibitem{bibTs:Boul} \emph{G. Bouligand}, \newblock Une forme donn\'{e}e a
la recherche des syst\`{e}mes triples orthogonaux, \newblock C. R. Acad.
Sci. Paris, v.236, p. 2462--2463 (1953).

\bibitem{Bogd+Fer} \emph{L. Bogdanov, E.V. Ferapontov}, \newblock A nonlocal
Hamiltonian formalism for semi-Hamiltonian systems of hydrodynamic type,
Theor. Math. Phys., \textbf{116} No. 1 (1998) 829--835.

\bibitem{BogdFer} \emph{L.~V. Bogdanov and E.~V. Ferapontov}, \newblock %
Projective differential geometry of higher reductions of the two-dimensional
Dirac equation.

\bibitem{bibTs:CEM} \emph{L. Chierchia, N. Ercolani, D. W. McLauhglin}, %
\newblock On the weak limit of rapidly oscillating waves, Duke Math. J., 
\textbf{55} (1987) 759--764.

\bibitem{bibTs:dar1} \emph{G. Darboux}, \newblock Le\c{c}ons sur les syst%
\`{e}mes orthogonaux et les coordonn\'{e}es curvilignes, Paris (1910).

\bibitem{bibTs:dar2} \emph{G. Darboux}, \newblock Le\c{c}ons sur la th\'{e}%
orie g\'{e}n\'{e}rale des surfaces et les applications g\'{e}om\'{e}triques
du calcul infinit\'{e}simal, t. 1--4, Paris (1887--1896).

\bibitem{bibTs:demoul} \emph{A. Demoulin}, \newblock Recherches sur les syst%
\`{e}mes triples orthogonaux, M\'{e}moirs Soc. Roy. Sciences Li\`{e}ge, v.
11, 98 p. (1921).

\bibitem{Dubr} \emph{B.A. Dubrovin,} \newblock Integrable systems in
topological field theory, Nucl. Phys. B, \textbf{379} (1992) 627--689. \emph{%
B.A. Dubrovin,} \newblock Hamiltonian formalism of Whitham-type hierarchies
and topological Landau-Ginsburg models, Comm. Math. Phys., \textbf{145}
(1992) 195--207. \emph{B.A. Dubrovin}, \newblock Geometry of 2D topological
field theories, Lecture Notes in Math. 1620, Springer-Verlag (1996)
120--348. \emph{B.A. Dubrovin}, \newblock On the differential geometry of
strongly integrable systems of hydrodynamic type, Funk. Anal. Appl., \textbf{%
24} (1990).

\bibitem{Dubr+Nov} \emph{B.~A. Dubrovin, S.~P. Novikov,} \newblock %
Hamiltonian formalism of one-dimensional systems of hydrodynamic type and
the Bogolyubov-Whitham averaging method, Soviet Math. Dokl., \textbf{27}
(1983) 781--785. \emph{B.~A. Dubrovin, S.~P. Novikov,} \newblock %
Hydrodynamics of weakly deformed soliton lattices. Differential geometry and
Hamiltonian theory, Russian Math. Surveys, \textbf{44} No. 6 (1989) 35--124.

\bibitem{bibTs:DzV} \emph{I.E. Dzyaloshinskii, G.E. Volovick}, \newblock %
Poisson brackets in condensed matter physics, Ann. of Phys.,~\textbf{125}
(1980) 67--97.

\bibitem{Egorov} \emph{D.F. Egorov}, \newblock Works in Differential
Geometry. Moscow, Nauka, 1970.

\bibitem{bibTs:Eisenh} \emph{L.P. Eisenhart}, \newblock Transformations of
surfaces, Princeton (1923), 2nd ed.- Chelsea (1962).

\bibitem{Erc} \emph{N. Ercolani et al.}, \newblock Hamiltonian structure for
the modulation equations of a sine-Gordon wavetrain, Duke Math. J., \textbf{%
55}~(1987) 949--983.

\bibitem{Fer+trans} \emph{E.~V. Ferapontov,} \newblock Nonlocal Hamiltonian
operators of hydrodynamic type: differential geometry and applications,
Amer. Math. Soc. Transl. (2), \textbf{170} (1995) 33--58. \emph{E.~V.
Ferapontov,} \newblock Differential geometry of nonlocal Hamiltonian
operators of hydrodynamic type, Func. Anal. Appl., \textbf{25} No. 3 (1991)
37--49.

\bibitem{viniti} \emph{E.V. Ferapontov}, \newblock Hamiltonian systems of
hydrodynamic type and their realization on hypersurfaces of a
pseudoeuclidean space, Geom. Sbornik, VINITI \textbf{22} (1990) 59--96
(English transl. in Soviet J. of Math., \textbf{55} (1991) 1970--1995).

\bibitem{Fer+comp} \emph{E.V. Ferapontov}, \newblock Compatible Poisson
brackets of hydrodynamic type, J. Phys. A: Math. Gen., \textbf{34} (2001)
1--12.

\bibitem{Maks+Fer} \emph{E.V. Ferapontov, M.V. Pavlov}, \newblock %
Quasiclassical limit of Coupled KdV equations. Riemann invariants and
multi-Hamiltonian structure. Physica D, \textbf{52} (1991), 211--219.

\bibitem{FerMaxNon} \emph{E.V. Ferapontov, M.V. Pavlov}, \newblock %
Reciprocal transformations of Hamiltonian operators of hydrodynamic type:
nonlocal Hamiltonian formalism for linearly degenerate systems, J. Math.
Phys., \textbf{44} No. 3 (2003) 1150--1172.

\bibitem{FFML} \emph{H. Flaschka, M.G. Forest, D.W. McLaughlin}, \newblock %
Multiphase averaging and the inverse spectral solution of the Korteweg-de
Vries equation, Comm. Pure Appl. Math., \textbf{33} (1980) 739.

\bibitem{Fordy+Gurel} \emph{A.P. Fordy, B. Gurel}, \newblock A new
construction of recursion operators for systems of hydrodynamic type. Theor.
Math. Phys., \textbf{122 }(2000) 37--49.

\bibitem{Ganzha} \emph{E. Ganzha}, \newblock Higher Lie-B\"{a}cklund
symmetries of diagonal systems of hydrodynamic type. (Russian)
Differentsialnye Uravneniya, \textbf{30} No. 10 (1994) 1725--1730;
translation in Diff. equations, \textbf{30} No. 10 (1994) 1593--1598. \emph{%
E. Ganzha}, \newblock Higher conservation laws of nonhomogeneous systems of
hydrodynamic type. (Russian) Vestnik Moskov. Univ. Ser. I Mat. Mekh., No. 1
(1996) 86--87; translation in Moscow Univ. Math. Bull., \textbf{51} No. 1
(1996) 46--47.

\bibitem{bibTs:Gui} \emph{C. Guichard}, \newblock Sur les syst\`{e}mes
triplement ind\'{e}termin\'{e}s et les syst\`{e}mes triples orthogonaux,
Collection Scientia, No. 25, Paris, 95 p. (1905). \emph{C. Guichard}, %
\newblock Th\'{e}orie des r\'{e}seaux, M\'{e}morial Sci. Math., fasc. 74,
Paris (1935). \emph{C. Guichard}, Th\'{e}orie g\'{e}n\'{e}rale des r\'{e}%
seaux, applications, M\'{e}morial Sci. Math., fasc. 77, Paris (1936).

\bibitem{bibTs:Haan} \emph{J. Haantjes}, \newblock On $X_{n-1}$-forming sets
of eigenvectors, Proc. Konink. Nederl. Akad. Wet., \textbf{58} No. 2 (1955);
Indagationes Mathematicae, \textbf{17} No. 2, 158--162.

\bibitem{Hay} \emph{W.D.~Hayes}, \newblock Group velocity and nonlinear wave
propagation, Proc. Roy. Soc. (London) Ser.~A, \textbf{332} (1973)~199--221.

\bibitem{bibTs:Hol} \emph{D.D.~Holm, B.A.~Kupershmidt}, \newblock Poison
brackets and Clebsch representation for magnetohydrodynamics, multifluid
plasmas, and electricity, Physica D, \textbf{6} (1983) 347--363. \emph{%
D.D.~Holm, B.A.~Kupershmidt}, \newblock Hamiltonian formulation of
ferromagnetic hydrodynamics, Phys. Letters A, \textbf{129} (1988) 93--100. 
\emph{D.D.~Holm}, \newblock Hamiltonian formulation of baroclinic
quasigeostrophic fluid equations, Phys. Fluids, \textbf{29} (1986) 7--8. 
\emph{D.D.~Holm}, \newblock Hamiltonian formulation of a charged fluid,
including electro- and magnetodynamics, Phys. Letters A, \textbf{114} (1986)
137--141. \emph{D.D.~Holm}, \newblock Hamiltonian formulation for Alfv\'{e}n
wave turbulence equations, Phys. Letters A, \textbf{108} (1985) 445--447.

\bibitem{bibTs:Katz} \emph{E.I. Katz, V.V. Lebedev}, \newblock Nonlinear
dynamics of smectic liquid chrystals with orientational ordering in the
layer, Soviet Phys. JETP, \textbf{61} (1985).

\bibitem{Krich} \emph{I.M. Krichever, }\newblock Algebraic-geometrical $N$
orthogonal curvilinear coordinate systems and solutions of the associativity
equations. Funct. Anal. Appl., \textbf{31} No. 1 (1997) 25--39.

\bibitem{Krich+Whitham} \emph{I.M. Krichever, }\newblock The averaging
method for two-dimensional ''integrable'' equations, Funct. Anal. Appl. 
\textbf{22} No. 3 (1988) 200--213. \emph{I.M. Krichever, }\newblock Spectral
theory of two-dimensional periodic operators and its applications, Russian
Math. Surveys \textbf{44} No. 2 (1989) 145--225.

\bibitem{Krich+wdvv} \emph{I.M. Krichever,} \newblock The dispersionless
equations and topological minimal models, Comm. Math. Phys., \textbf{143 }%
No. 2 (1992) 415--429. \emph{I.M. Krichever,} \newblock The $\tau $-function
of the universal Whitham hierarchy, matrix models and topological field
theories, Comm. Pure Appl. Math., \textbf{47} (1994) 437--475. \emph{A.A.
Akhmetshin, I.M. Krichever, Y.S. Volvovski,} \newblock A generating formula
for the solutions of the associativity equations. Russian Math. Surveys, 
\textbf{54} No. 2 (1999) 427--429.

\bibitem{Lax} \emph{Lax~P.D,} \newblock Hyperbolic systems of conservation
laws II. Comm. Pure Appl. Math., \textbf{10} (1957) 537--566. \emph{P.D.
Lax, C.D. Levermore}, \newblock The Zero Dispersion Limit for the
Korteweg-de Vries Equation, Proc. Nat. Acad. Sci. USA, \textbf{76}(8) (1979)
3602--3606. \emph{P.D. Lax and C.D. Levermore}, \newblock The Small
Dispersion Limit for the Korteweg-de Vries Equation I, II, III, Comm. Pure
Appl. Math., \textbf{36} (1983) 253--290, 571--593, 809--830.

\bibitem{Magri} \emph{F. Magri}, \newblock A simple model of the integrable
Hamiltonian system, J. Math. Phys., \textbf{19} (1978) 1156--1162.

\bibitem{Malt+Nov} \emph{A.Ya. Maltsev, S.P. Novikov}, \newblock On the
local systems Hamiltonian in the weakly nonlocal Poisson brackets, Physica
D, \textbf{156} (2001) 53--80.

\bibitem{Mokh} \emph{O.I. Mokhov}, \newblock Nonlocal Hamiltonian operators
of hydrodynamic type with flat metrics, integrable hierarchies and the
associativity equation. (Russian) Funk. Anal. i Pril., \textbf{40} No. 1
(2006) 14--29; translation in Funct. Anal. Appl., \textbf{40} No. 1 (2006)
11--23.

\bibitem{Mokh+comp} \emph{O.I. Mokhov}, \newblock Compatible flat metrics.
J. Appl. Math., \textbf{2} No. 7 (2002) 337--370. \emph{O.I. Mokhov}, %
\newblock On the integrability of equations for nonsingular pairs of
compatible flat metrics. (Russian) Teoret. Mat. Fiz., \textbf{130} No. 2
(2002) 233--250; translation in Theoret. and Math. Phys., \textbf{130} No. 2
(2002) 198--212.

\bibitem{Mokhov} \emph{O.I. Mokhov}, \newblock Symplectic forms on a loop
space and Riemannian geometry. (Russian) Funk. Anal. i Pril., \textbf{24}
No. 3 (1990) 86--87; translation in Funct. Anal. Appl., \textbf{24} No. 3
(1990) 247--249.

\bibitem{Nutku} \emph{Y. Nutku}, \newblock On a new class of completely
integrable nonlinear wave equations. Multi-Hamiltonian structure II, J.
Math. Phys., \textbf{28} No. 11 (1987) 2579--2585.

\bibitem{NO} \emph{Ya. Nutku, P. Olver}, \newblock Hamiltonian structures
for systems of hyperbolic conservation laws, J. Math. Phys., \textbf{29}
(1988) 1610--1619.

\bibitem{Lagrang} \emph{Ya. Nutku, M.V. Pavlov}, \newblock Multi-Lagrangians
for Integrable Systems, J. Math. Phys., \textbf{43} No. 3 (2002) 1441--1459.

\bibitem{Maks+multi} \emph{M.V. Pavlov,} \newblock Elliptic coordinates and
multi-Hamiltonian structures of hydrodynamic type systems. Russian Acad.
Sci. Dokl. Math., \textbf{50} No. 3 (1995) 374--377. \emph{M.V. Pavlov,} %
\newblock Multi-Hamiltonian structures of the Whitham equations. Russian
Acad. Sci. Dokl. Math., \textbf{50} No. 2 (1995) 220--223.

\bibitem{Maks+int} \emph{M.V. Pavlov}, \newblock Integrability of the Egorov
hydrodynamic type systems. submitted.

\bibitem{Maks+wdvv} \emph{M.V. Pavlov}, \newblock Explicit solutions of the
WDVV equation determined by the ``flat'' hydrodynamic reductions of the
Egorov hydrodynamic chains. submitted.

\bibitem{Maks+comp} \emph{M. V. Pavlov,} \newblock Description of compatible
pairs of differential-geometric Poisson brackets of the first order. Theor.
Math. Phys., \textbf{142} No. 2 (2005) 293--309.

\bibitem{Maks+algebr} \emph{M. V. Pavlov,} \newblock Algebro-geometric
approach in the theory of integrable hydrodynamic type systems. submitted.

\bibitem{Maks+Tsar} \emph{M. V. Pavlov, S.P. Tsarev,} \newblock
Three-Hamiltonian structures of the Egorov hydrodynamic type systems, Funct.
Anal. Appl., \textbf{37} No. 1 (2003) 32--45.

\bibitem{Yanenko} \emph{B.L. Rozhdestvenski, N.N. Yanenko}, \newblock %
Systems of quasilinear equations and their applications to gas dynamics.
Translated from the second Russian edition by J. R. Schulenberger.
Translations of Mathematical Monographs, 55. American Mathematical Society,
Providence, RI, 1983; Russian ed. Nauka, (1968) Moscow.

\bibitem{Serre} \emph{D. Serre}, \newblock Invariants du premier ordre des
systemes hyperboliques semi-lineares. J. of Diff. Eq., \textbf{49}
(1983)~270--280. \emph{D. Serre}, \newblock Syst\`{e}mes d'EDO invariants
sous l'action de syst\`{e}mes hyperboliques d'EDP, Ann. Inst. Fourier, 
\textbf{39} (1989) 953. \emph{D. Serre}, \newblock Int'{e}grabilit'{e} d'une
classe de syst\`{e}mes de lois de conservation, \newblock Preprint ENS Lyon, 
\textbf{45} (1991).

\bibitem{Sheftel} \emph{M.B. Sheftel}, \newblock On the Lie-B\"{a}cklund
groups admissible by the equations of one-dimensional gas dynamics, Vestnik
of Leningrad State Univ. No. 7 (1982) 37--41 (in Russian). \emph{M.B. Sheftel%
}, \newblock On the infinite-dimensional noncommutative Lie-B\"{a}cklund
algebra associated with the equations of one-dimensional gas dynamics,
Theor. Math. Phys., \textbf{56} (1983) 878--891. \emph{M.B. Sheftel}, %
\newblock Integration of Hamiltonian systems of hydrodynamic type with two
dependent variables with the aid of the Lie-B\"{a}cklund group, Func. Anal.
Appl., \textbf{20} (1986)~227--235. \emph{M.B. Sheftel}, \newblock Higher
integrals and symmetries of semi-Hamiltonian systems, Diff. equations, 
\textbf{29} No.10 (1993)~1782--1795.

\bibitem{Tenenblat} \emph{K. Tenenblat}, \newblock B\"{a}cklund's theorem
for submanifolds of space forms and a generalized wave equation, Boll. Soc.
Brasil. Mat. \textbf{16} (1985) 67--92. \emph{K. Tenenblat, C.L. Terng}, %
\newblock B\"{a}cklund's theorem for $N$-dimensional submanifolds of $%
R^{2N-1}$, Ann. Math. \textbf{111} (1980) 477--490. \emph{C.L. Terng}, %
\newblock Soliton equations and differential geometry, J. Diff. Geom., 
\textbf{45} (1997) 407--445. \emph{M. Dajczer, R. Tojeiro}, \newblock An
extension of the classical Ribaucour transformation, Proc. London Math. Soc. 
\textbf{85} (2002) 211-232.

\bibitem{Terng} \emph{Chuu-Lian Terng}, \newblock Geometries and symmetries
of solition equations and integrable elliptic equations (2002). \emph{%
Chuu-Lian Terng, Karen Uhlenbeck}, \newblock Poisson actions and scattering
theory for integrable systems (1997).

\bibitem{Teshukov} \emph{V.M. Teshukov}, \newblock Hyperbolic systems
admitting a nontrivial Lie-Backlund group, \newblock Preprint LIIAN \textbf{%
106} (1989) 25--30.

\bibitem{Tsar} \emph{S.P. Tsarev}, \newblock On Poisson brackets and
one-dimensional Hamiltonian systems of hydrodynamic type, Soviet Math.
Dokl., \textbf{31} (1985) 488--491. \emph{S.P. Tsarev}, \newblock The
geometry of Hamiltonian systems of hydrodynamic type. The generalized
hodograph method, Math. USSR Izvestiya, \textbf{37} No. 2 (1991) 397--419. 
\emph{S.P. Tsarev}, \newblock On the integrability of the averaged KdV and
Benney equations, Proc. NATO ARW ``Singular Limits and Dispersive Waves'',
Lyon, France, 8--12 July, 1991. Ercolani, N. M. (ed.) et al., New York, NY:
Plenum, NATO ASI Ser., Ser. B, Phys. 320, 143--155 (1994).

\bibitem{Tsar+geom} \emph{S.P. Tsarev}, \newblock Classical differential
geometry and integrability of systems of hydrodynamic type. In: Proc. NATO
ARW ``Applications of Analytic and Geometrical Methods to Nonlinear
Differential Equations'', 14--19 July, 1992, Exeter, UK, NATO ASI Series C
413, (ed. P.A.Clarkson), Kluwer Publ., p. 241--249. \emph{S.P. Tsarev}, %
\newblock Integrability of equations of hydrodynamic type from the end of
the 19th to the end of the 20th century. In: ``Integrability: the
Seiberg-Witten and Whitham equations'' (Edinburgh, 1998) p. 251--265, Gordon
and Breach, Amsterdam, 2000.

\bibitem{Veroski} \emph{J.M. Verosky}, \newblock Higher-order symmetries of
the compressible one-dimensional isentropic fluid equations, J. Math. Phys., 
\textbf{25} (1984) 884. \emph{J.M. Verosky}, \newblock First-order conserved
densities for gas dynamics, J. Math. Phys. \textbf{27,} No. 12 (1986)
3061--3063.

\bibitem{Zakh} \emph{V.E. Zakharov}, \newblock Description of the $N$
orthogonal curvilinear coordinate systems and Hamiltonian integrable systems
of hydrodynamic type, I: Integration of the Lame equations. Duke Math. J. 
\textbf{94} No. 1 (1998) 103--139. \emph{V.E. Zakharov}, \newblock %
Integration of the Gauss-Codazzi equations, Teor. Math. Phys., \textbf{128}
No. 1 (2001) 946--956. \emph{V.E. Zakharov}, \newblock Application of the
Inverse Scattering Transform to Classical Problems of Differential Geometry
and General Relativity, Contemporary Mathematics, \textbf{301} (2002).
\end{thebibliography}
\end{document}